%% file: EXO-16-050_temp.tex
\pdfoutput=1

\documentclass[11pt,twoside,a4paper,cmspaper,final,collab]{cms-tdr}
\def\svnVersion{493285P}\def\cmsCernNoTag{CERN-EP-2018-287}\def\cmsCernDate{\today}\def\cmsMessage{Published in the European Physical Journal C as \href{http://dx.doi.org/10.1140/epjc/s10052-019-6730-7}{\doi{10.1140/epjc/s10052-019-6730-7.}}}

\begin{document}\cmsNoteHeader{EXO-16-050}

\hyphenation{had-ron-i-za-tion}
\hyphenation{cal-or-i-me-ter}
\hyphenation{de-vices}
\RCS$HeadURL: svn+ssh://svn.cern.ch/reps/tdr2/papers/EXO-16-050/trunk/EXO-16-050.tex $
\RCS$Id: EXO-16-050.tex 464952 2018-06-18 09:24:35Z bmaier $
\newlength\cmsFigWidth
\ifthenelse{\boolean{cms@external}}{\setlength\cmsFigWidth{0.85\columnwidth}}{\setlength\cmsFigWidth{0.4\textwidth}}
\ifthenelse{\boolean{cms@external}}{\providecommand{\cmsLeft}{top\xspace}}{\providecommand{\cmsLeft}{left\xspace}}
\ifthenelse{\boolean{cms@external}}{\providecommand{\cmsRight}{bottom\xspace}}{\providecommand{\cmsRight}{right\xspace}}

\newcommand{\SigSI}{\ensuremath{\sigma_\mathrm{SI}}\xspace}

\newlength\cmsTabSkip\setlength{\cmsTabSkip}{1ex}

\ifthenelse{\boolean{cms@external}}{\providecommand{\cmsTable}[1]{#1}}{\providecommand{\cmsTable}[1]{\resizebox{\textwidth}{!}{#1}}}

\cmsNoteHeader{EXO-16-050}
\title{Search for dark matter produced in association with a Higgs boson decaying to a pair of bottom quarks in proton-proton collisions at $\sqrt{s}=13\TeV$}

\date{\today}

\abstract{A search for dark matter produced in association with a Higgs boson decaying to a pair of bottom quarks is performed in
 proton-proton collisions at a center-of-mass energy of 13\TeV collected with the CMS detector at the LHC. The analyzed data sample corresponds to an integrated luminosity of 35.9\fbinv.
The signal is characterized by a large missing transverse momentum recoiling against a bottom quark-antiquark system that has a large Lorentz boost. The number of events observed in the data is consistent with the standard model background prediction.  Results are interpreted in terms of limits both on parameters of the type-2 two-Higgs doublet model extended by an additional light pseudoscalar boson \Pa (2HDM+\Pa) and on parameters of a baryonic $\cPZpr$ simplified model. The 2HDM+\Pa model is tested experimentally for the first time. For the baryonic $\cPZpr$ model, the presented results constitute the most stringent constraints to date. 
}

\hypersetup{%
pdfauthor={CMS Collaboration},%
pdftitle={Search for dark matter produced in association with a Higgs boson decaying to a pair of bottom quarks in proton-proton collisions at sqrt(s)=13 TeV},%
pdfsubject={CMS},%
pdfkeywords={Dark matter, LHC, CMS, boosted Higgs boson tagging}}

\maketitle

\section{Introduction} \label{intro}

Astrophysical evidence for dark matter (DM) is one of the most compelling motivations for physics beyond the standard model (SM)~\cite{dm1,dm2,dm3}.
Cosmological observations demonstrate that around 85\% of the matter in the universe is comprised of DM \cite{planck} and they are largely consistent with the hypothesis that DM is  composed primarily of weakly interacting massive particles.
If nongravitational interactions exist between DM and SM particles, DM could be produced by colliding SM particles at high energy.
Assuming the pair production of DM particles in hadron collisions occurs through a spin-0 or spin-1 bosonic mediator coupled to the initial-state particles, the DM particles leave the detector without  measurable signatures.
If DM particles are produced in association with a detectable SM particle, which could be emitted as initial-state radiation from the interacting constituents of the colliding protons, or through new effective couplings between DM and SM particles, their existence could be inferred via a large transverse momentum imbalance in the collision event.

The production of the SM Higgs boson~\cite{HiggsObs_ATLAS, HiggsObs_CMS, HiggsObs_CMS_Long} via initial-state radiation is highly suppressed because of the mass dependence of its coupling strength to fermions.
Nonetheless, the associated production of a Higgs boson and DM particles can occur if the Higgs boson takes part in the interaction producing the DM particles~\cite{monoHiggs3,2HDM,PhysRevD.89.075017}.
Such a production mechanism would allow one to directly probe the structure of the effective DM-SM coupling.

\begin{figure*}[t!]
\centering
 \includegraphics[width=0.4\textwidth]{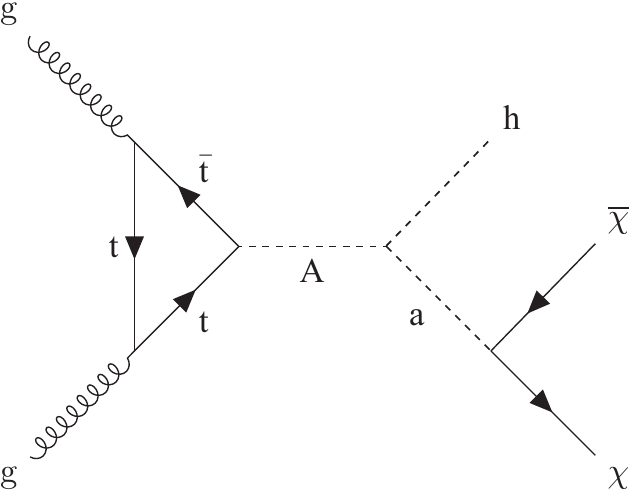}\hspace{1cm}
 \includegraphics[width=0.34\textwidth]{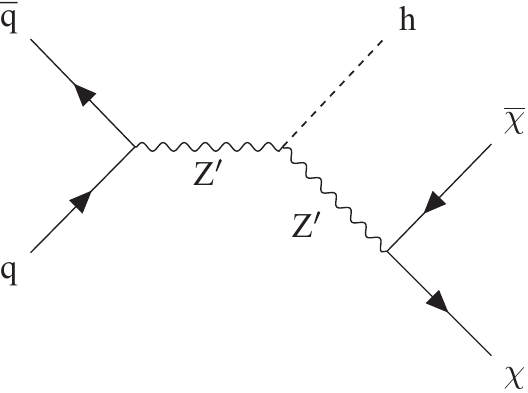} \\
\caption{Feynman diagrams for the 2HDM+\Pa model (left) and the baryonic \cPZpr\ model (right). In both models, the scalar \Ph can be identified with the observed 125\GeV Higgs boson.}
\label{feyns}
\end{figure*}

In this paper, we present a search for DM production in association with a scalar Higgs boson, h, with a mass of 125\GeV that decays to a bottom quark-antiquark pair (\bbbar).
As the \bbbar decay mode has the largest branching fraction of all Higgs boson decay modes allowed in the SM, it provides the largest signal yield.
The search is performed using the data set collected by the CMS experiment~\cite{CMSdetector} at the CERN Large Hadron Collider (LHC) at a center-of-mass energy of 13\TeV in 2016, corresponding to an integrated luminosity of approximately 35.9\fbinv.
Similar searches have been conducted at the LHC by both the ATLAS and CMS Collaborations, analyzing data collected at 8~\cite{PhysRevLett.115.131801} and 13\TeV~\cite{PhysRevLett.119.181804,1807.02826}.
Results are interpreted in terms of two simplified models predicting this signature.
The first is a type-2 two-Higgs doublet model extended by an additional light pseudoscalar boson \Pa (2HDM+\Pa)~\cite{Bauer2017}.
The \Pa boson mixes with the scalar and pseudoscalar partners of the observed Higgs boson, and decays to a pair of DM particles,  $\chi\overline{\chi}$.
The second is a baryonic $\cPZpr$ model~\cite{PhysRevD.89.075017}, in which a ``baryonic Higgs'' boson mixes with the SM Higgs boson. In this model, a vector mediator $\cPZpr$ is exchanged in the $s$-channel and, after the radiation of an SM Higgs boson, decays to two DM particles.
Representative Feynman diagrams for the two models are presented in Fig.~\ref{feyns}.

In the 2HDM+\Pa model, the DM particle candidate $\chi$ is a fermion that can couple to SM particles only through a spin-0, pseudoscalar mediator.
Since the couplings of the new spin-0 mediator to SM gauge bosons are strongly suppressed, the 2HDM+\Pa model is consistent with measurements of the SM Higgs boson production and decay modes, which so far show no significant deviation from SM predictions~\cite{Khachatryan:2016vau}.
In contrast to previously explored two-Higgs doublet models~\cite{2HDM,PhysRevLett.115.131801,PhysRevLett.119.181804,Sirunyan:2017hnk}, the 2HDM+\Pa framework ensures gauge invariance and renormalizability.
In this model there are six mass eigenstates.
Two are charge-parity (CP)-even scalars: the light \Ph, assumed to be the observed 125\GeV Higgs boson, and the heavy \PH.
These are the result of the mixing of the neutral CP-even weak eigenstates with a mixing angle $\alpha$.
The two CP-odd pseudoscalar mass eigenstates are the light \Pa and the heavy \PSA, which are linear combinations of the CP-odd weak eigenstates, with a mixing angle $\theta$.
Finally, there are two heavy charged scalars \Hpm with identical mass.

The masses of \Pa and \PSA, the angle $\theta$, and the ratio of the vacuum expectation values of \Ph and \PH, $\tan\beta$, are varied in this search. 
The mixing angle $\alpha$ changes with $\beta$ following the relation $\alpha = \beta - \pi/2$.
Perturbativity and unitarity put restrictions on the magnitudes and the signs of the three quartic couplings $\lambda_3,~\lambda_{\mathrm{P}1},~\lambda_{\mathrm{P}2}$, and we set their values to $\lambda_3=\lambda_{\mathrm{P}1}=\lambda_{\mathrm{P}2}=3$~\cite{Bauer2017}.
The masses of the charged Higgs bosons and of the heavy CP-even Higgs boson are assumed to be the same as the mass of the heavy pseudoscalar, \ie, $m_{\PH}=m_{\Hpm}=m_\PSA$.
When performing a scan in the $m_\PSA$-$m_\Pa$ plane, $\tan\beta$ is assumed to be 1 and $\sin\theta$ is assumed to be 0.35, following the recommendations in Ref.~\cite{Abe:2018bpo}.
The DM particle $\chi$ is assumed to have a mass of $m_\chi=10\GeV$.
For $\tan\beta\gg1$, the coupling strengths of both \Pa and \PSA to {\cPqb} quarks are enhanced, and effects from \bbbar-initiated production are included in the signal simulation for all values of $\tan\beta$.

The baryonic \cPZpr\ model~\cite{PhysRevD.89.075017} is an extension of the SM with an additional U(1)$_{B}$ \cPZpr\ gauge boson that couples to the baryon number $B$.
The model predicts the existence of a new Dirac fermion that is neutral under SM gauge symmetries, has non-zero $B$, and is stable because of the corresponding U(1)$_{B}$ symmetry.
The state therefore serves as a good DM candidate.
To generate the  \cPZpr\ mass, a baryonic Higgs scalar field is introduced to spontaneously break the U(1)$_B$ symmetry.
In analogy with the SM, there remains a physical baryonic Higgs particle, $\Ph_{B}$, with a vacuum expectation value $v_{B}$, which couples to the \cPZpr\ boson.
The \cPZpr\ and the SM Higgs boson, \Ph, interact with a coupling strength of $g_{\Ph\cPZpr\cPZpr}$ = $m_{\cPZpr}^{2} \sin \zeta/v_{B}$, where $\zeta$ is the \Ph-$\Ph_{B}$ mixing angle.
The chosen value for the \cPZpr\ coupling to quarks, $g_\cPq$, is 0.25 and the \cPZpr\ coupling to DM, $g_\chi$, is set to 1, following the recommendations in Ref.~\cite{Abercrombie:2015wmb}.
This is well below the bounds $g_\cPq,g_\chi\sim4\pi$, where perturbativity and the validity of the effective field theory break down~\cite{PhysRevD.89.075017}.
Constraints on the SM Higgs boson properties make the mixing angle $\zeta$ consistent with $\cos\zeta =1$ within uncertainties of the order of 10\%, thereby requiring  $\sin\zeta$ to be less than 0.4~\cite{PhysRevD.89.075017}.
In this search, it is assumed that $\sin\zeta= 0.3$.
It is also assumed that $g_{\Ph\cPZpr\cPZpr}/m_{\cPZpr}=1$, which implies $v_B=m_{\cPZpr}\sin\zeta$.
This choice maximizes the cross section without violating the bounds imposed by SM measurements.
The free parameters in the model under these assumptions are thus $m_{\cPZpr}$ and $m_\chi$, which are varied in this search.

Signal events are characterized by a large imbalance in the transverse momentum (or hadronic recoil), which indicates the presence of invisible DM particles, and by hadronic activity consistent with the production of an SM Higgs boson that decays to a \bbbar pair.
Thus, the search strategy followed imposes requirements on the mass of the reconstructed Higgs boson candidate, which is also required to be Lorentz-boosted.
Together with the identification of the hadronization products of the two \cPqb~quarks produced in the Higgs boson decay, these requirements define a data sample that is expected to be enriched in signal events.
Several different SM processes can mimic this topology, the most important of which are top quark pair production and the production of a vector boson (V) in association with multiple jets.
For each of these SM processes that constitute the largest sources of background, statistically independent data samples are used to predict the hadronic recoil distributions.
Both the signal and background contributions to the hadronic recoil distributions observed in data are extracted with a likelihood fit, performed simultaneously in all samples.

\section{The CMS detector}

The CMS detector, described in detail in Ref.~\cite{CMSdetector}, is a multipurpose apparatus designed to study high transverse momentum (\pt) processes in proton-proton (pp) and heavy ion collisions.
A superconducting solenoid occupies its central region, providing a magnetic field of 3.8\unit{T} parallel to the beam direction.
Charged particle trajectories are measured using silicon pixel and strip trackers that cover a pseudorapidity region of $\abs{\eta} < 2.5$.
A lead tungstate crystal electromagnetic calorimeter (ECAL), and a brass and scintillator hadron calorimeter  surround the tracking volume and extend to $\abs{\eta} < 3$.
The steel and quartz-fiber forward Cherenkov hadron calorimeter extends the coverage to $\abs{\eta} < 5$.
The muon system consists of gas-ionization detectors embedded in the steel flux-return yoke outside the solenoid and covers $\abs{\eta} < 2.4$.
Online event selection is accomplished via the two-tiered CMS trigger system~\cite{1748-0221-12-01-P01020}. The first level is designed to select events in less than 4\mus, using information from the calorimeters and muon detectors.
Subsequently, the high-level trigger processor farm reduces the event rate to 1\,kHz.

\section{Simulated data samples}

The signal processes are simulated at leading order (LO) accuracy in quantum chromodynamics (QCD) perturbation theory using the \MGvATNLO v2.4.2~\cite{amcatnlo} program.
To model the contributions from SM Higgs boson processes as well as from the \ttbar and single top quark backgrounds, we use the {\POWHEG~v2}~\cite{Nason:2004rx,Frixione:2007vw,Alioli:2010xd} generator.
These processes are generated at the next-to-leading order (NLO) in QCD.
The \ttbar production cross section is further corrected using calculations at the next-to-next-to-leading order in QCD including corrections for soft-gluon radiation estimated with next-to-next-to-leading logarithmic accuracy~\cite{ttbarNNLO}.
Events with multiple jets produced via the strong interaction (referred to as QCD multijet events) are generated at LO using \MGvATNLO v2.2.2 with up to four partons in the matrix element calculations.
The MLM prescription~\cite{mlm} is used for matching these partons to parton shower jets.
Simulated samples of {\cPZ}+jets and {\PW}+jets processes are generated at LO using \MGvATNLO v2.3.3. Up to four additional partons are considered in the matrix element and matched to their parton showers using the MLM technique.
The V+jets (V={\PW},{\cPZ}) samples are corrected by weighting the \pt of the respective boson with NLO QCD corrections obtained from large samples of events generated with \MGvATNLO and the FxFx merging technique~\cite{fxfx} with up to two additional jets stemming from the matrix element calculations.
These samples are further corrected by applying NLO electroweak corrections~\cite{Kuhn:2005gv,Kallweit:2015fta,Kallweit:2015dum} that depend on the boson \pt.
Predictions for the SM diboson production modes {\PW}{\PW}, {\PW}{\cPZ}, and {\cPZ}{\cPZ} are obtained at LO with the {\PYTHIA 8.205}~\cite{Sjostrand:2014zea} generator and normalized to NLO accuracy using {\MCFM v6.0}~\cite{MCFM}.

The LO or NLO NNPDF 3.0 parton distribution functions (PDFs)~\cite{Ball:2014uwa} are used, depending on the QCD order of the generator used for each physics process.
Parton showering, fragmentation, and hadronization are simulated with {\PYTHIA 8.212} using the CUETP8M1 underlying event tune~\cite{ue1,ue2}.
Interactions of the resulting final state particles with the CMS detector are simulated using the \GEANTfour program~\cite{geant4}.
Additional inelastic pp interactions in the same or a neighboring bunch crossing (pileup) are included in the simulation.
The pileup distribution is adjusted to match the corresponding distribution observed in data.

\section{Event reconstruction}

The reconstructed interaction vertex with the largest value of summed physics-object $\pt^2$ is taken to be the primary event vertex.
The physics objects used for the primary event vertex determination are the clusters found by the anti-$\kt$ clustering algorithm~\cite{Cacciari:2008gp,Cacciari:2011ma}, with a distance parameter of 0.4, from the charged particle tracks in the event, as well as the associated missing transverse momentum, taken as the negative vector sum of the \pt of those clusters.
The offline selection requires all events to have a primary vertex reconstructed within a 24\cm window along the $z$-axis around the nominal interaction point, and a transverse distance from the nominal interaction region less than 2\cm.

The particle-flow (PF) algorithm~\cite{Sirunyan:2017ulk} aims to reconstruct and identify each individual particle in an event, with an optimized combination of information from the various elements of the CMS detector.
The energy of photons is obtained from the ECAL measurement.
The energy of electrons is determined from a combination of the electron momentum at the primary interaction vertex as determined by the tracker, the energy of the corresponding ECAL cluster, and the energy sum of all bremsstrahlung photons spatially compatible with originating from the electron track.
The energy of muons is obtained from the curvature of the corresponding track. The energy of charged hadrons is determined from a combination of their momentum measured in the tracker and the matching ECAL and HCAL energy deposits, corrected for zero-suppression effects and for the response function of the calorimeters to hadronic showers.
Finally, the energy of neutral hadrons is obtained from the corresponding corrected ECAL and HCAL energies.
The PF candidates are then used to construct the physics objects described in this section.
At large Lorentz boosts, the two {\cPqb} quarks from the Higgs boson decay may produce jets that overlap and make their individual reconstruction difficult.
In this search large-area jets clustered from PF candidates using the Cambridge--Aachen algorithm~\cite{cajets} with a distance parameter of 1.5 (CA15 jets) are utilized to identify the Higgs boson candidate.
The large cone size is chosen in order to select signal events where the Higgs boson has a medium Lorentz-boost and hence its decay products begin to merge for $\pt(\Ph)\gtrsim200\GeV$.
To reduce the impact of particles arising from pileup interactions,  the four-vector of each PF candidate is scaled with a weight calculated with the pileup per particle identification (PUPPI) algorithm~\cite{puppi} prior to the clustering.
The absolute jet energy scale is corrected using calibrations derived from data~\cite{jec}.
The CA15 jets are also required to be central ($\abs{\eta} < 2.4$).
The ``soft-drop'' (SD) jet grooming algorithm~\cite{msd} is applied to remove soft wide-angle radiation from the jets.
We refer to the mass of the groomed CA15 jet as $m_\mathrm{SD}$.

The ability to identify two {\cPqb} quarks inside a single CA15 jet is crucial for this search.
A likelihood for the CA15 jet to contain two {\cPqb} quarks is derived by combining the information from primary and secondary vertices and tracks in a multivariate discriminant optimized to distinguish CA15 jets originating from $\Ph\to \bbbar$ decays from the cases where the hadronization of energetic quarks or gluons~\cite{Sirunyan:2017ezt} leads to the presence of a CA15 jet.
The working point chosen for this algorithm (the ``double-{\cPqb} tagger'') corresponds to an identification efficiency of 50\% for a \bbbar system with a \pt of 200\GeV, and a probability of 10\% for misidentifying CA15 jets originating from combinations of quarks or gluons not coming from a resonance decay. The efficiency of the algorithm increases with the \pt of the \bbbar system, reaching an efficiency of 65\% for a CA15 jet with a $\pt>500\GeV$. In this \pt regime, the misidentification rate for QCD jets is about 13\%. The probability for misidentifying CA15 jets from top quark decays is 14\% across the entire \pt spectrum. These estimates are derived with no additional requirements on the CA15 jet kinematics.

Energy correlation functions are used to identify the two-prong structure in the CA15 jet expected from a Higgs boson decay to two \cPqb~quarks, and to distinguish it from QCD-like jets (\ie, jets that do not originate from a heavy resonance decay) and jets from the hadronic decays of top quarks.
The energy correlation functions (${_ve}_N$) are sensitive to correlations among the constituents of the CA15 jet~\cite{ecf} and depend on $N$ factors of the particle energies and $v$ factors of their pairwise angles, weighted by the angular separation of the constituents.

As motivated in Ref.~\cite{ecf}, the ratio $N_2 = {_2e}_3^{(\beta)}/(_1e_2^{(\beta)})^2$ is used as a two-prong tagger for the identification of the CA15 jet containing the Higgs boson decay products.
The parameter $\beta$, which controls the weighting of the angles between constituent pairs in the computation of the $N_2$ variable, is chosen to be 1 since this value gives the best two-prong jet identification.

\begin{figure}
\centering
 \includegraphics[width=0.475\textwidth]{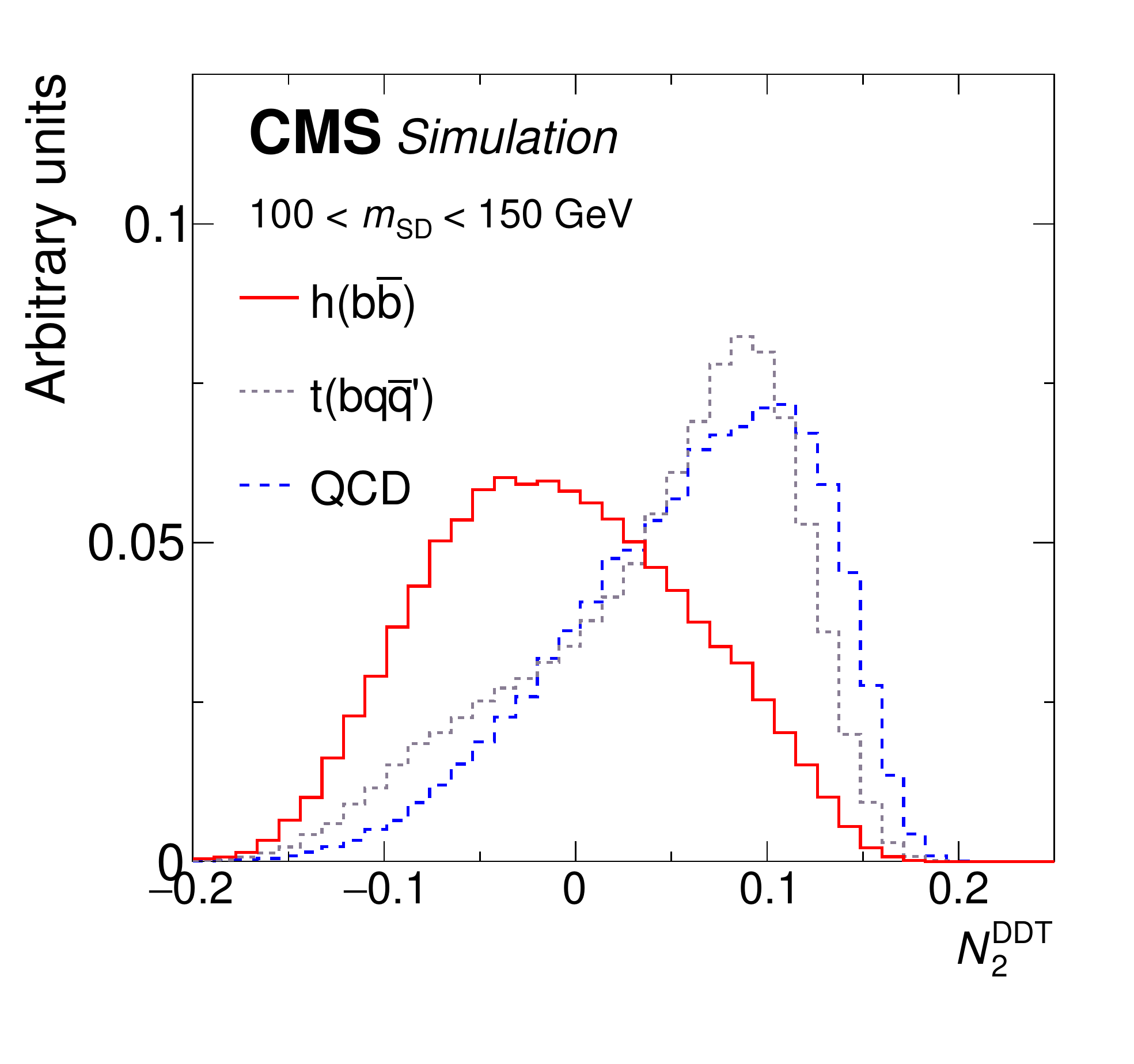}
 \caption{The $N_2^\text{DDT}$ distribution as expected for CA15 jets originating from a Higgs boson decaying to a \bbbar pair (solid red) is compared with the expected distribution for CA15 jets originating from the decay products of top quarks decaying hadronically (dotted grey). The distribution corresponding to CA15 jets that do not originate from a heavy resonance decay is also shown (dashed blue).}
\label{n2ddt}
\end{figure}

It is noted that requiring a jet to be two-pronged based on the value of a jet substructure variable, such as $N_2$, will affect the shape of the distribution of $m_\mathrm{SD}$ for the background processes.
In this search, the value of $m_\mathrm{SD}$ is required to be consistent with the Higgs boson mass.
It is therefore desirable to preserve a smoothly falling jet mass distribution for QCD-like jets.
As motivated in Ref.~\cite{ddt}, the dependence of $N_2$ on the variable $\rho=\ln(m_{\mathrm{SD}}^2/\pt^2)$ is tested, since the distribution of $\rho$ in QCD-like jets is expected to be invariant of the jet mass and \pt.
The decorrelation strategy described in Ref.~\cite{ddt} is applied, choosing a QCD misidentification efficiency of 20\%, which corresponds to a signal efficiency of 55\% and a misidentification rate for top quark jets of 36\% across the entire CA15 jet \pt spectrum.
This results in a modified tagging variable, which we denote as $N_2^\mathrm{DDT}$, where the superscript DDT stands for ``designing decorrelated taggers''~\cite{ddt}.
Figure~\ref{n2ddt} shows the expected distribution of $N_2^\mathrm{DDT}$ for CA15 jets matched to a Higgs boson decaying to a \bbbar pair, together with the distributions expected for CA15 jets matched to hadronically decaying top quarks and for QCD-like CA15 jets.

This search also utilizes narrow jets clustered from the PF candidates using the anti-$\kt$ algorithm with a distance parameter of 0.4 (``AK4 jets'').
Narrow jets originating from {\cPqb} quarks are identified using the combined secondary vertex (CSVv2) algorithm \cite{Sirunyan:2017ezt}.
The working point used in this search has a {\cPqb}-jet identification efficiency of 81\%, a charm jet selection efficiency of 37\%, and a 9\% probability of misidentifying light-flavor jets~\cite{Sirunyan:2017ezt}.
Jets that are {\cPqb}-tagged are required to be central ($\abs{\eta}<2.4$).

Electron reconstruction requires the matching of a supercluster in the ECAL with a track in the silicon tracker.
Reconstructed electrons are required to be within $\abs{\eta}< 2.5$, excluding the transition region $1.44<\abs{\eta}<1.57$ between the ECAL barrel and endcap.
Identification criteria~\cite{Khachatryan:2015hwa} based on the ECAL shower shape and the consistency of the electron track with the primary vertex are imposed.
Muon candidates are selected by two different reconstruction approaches~\cite{CMSMuonJINST}: one in which tracks in the silicon tracker are matched to a track segment in the muon detector, and another in which a track fit spanning the silicon tracker and muon detector is performed starting with track segments in the muon detector.
Further identification criteria are imposed on muon candidates to reduce the number of hadrons and poorly measured mesons misidentified as muons~\cite{CMSMuonJINST}.
These additional criteria include requirements on the number of hits in the tracker and in the muon systems, the fit quality of the global muon track, and the track's consistency with the primary vertex.
Muon candidates with $\abs{\eta}< 2.4$ are considered in this analysis.
A minimum \pt of 10\GeV is required for electron and muon candidates.
Both are required to satisfy isolation requirements that limit the total energy of tracks and calorimeter clusters measured in conical regions about them.
Hadronically decaying $\tau$ leptons, $\tau_\text{had}$, are reconstructed using the hadron-plus-strips  algorithm~\cite{CMSTauJINST}, which uses charged hadron and neutral electromagnetic objects  to reconstruct intermediate resonances into which the $\tau$ lepton decays.
The $\tau_\text{had}$ candidates with $\pt>18\GeV$ and $\abs{\eta}< 2.3$ are considered~\cite{Khachatryan:2015hwa,Chatrchyan:2013sba,CMSTauJINST}.
Photon candidates, identified by means of requirements on the ECAL energy distribution and its distance to the closest track, must have $\pt>15\GeV$ and $\abs{\eta}< 2.5$.

The missing transverse momentum \ptvecmiss is defined as the negative vectorial sum of the \pt of all the reconstructed PF candidates.
Its magnitude is denoted as \ptmiss.
Corrections to jet momenta are propagated to the \ptmiss calculation, and event filters are used to remove spurious high \ptmiss events caused by instrumental noise in the calorimeters or beam halo muons~\cite{CMS-PAS-JME-16-004}.
These filters remove about 1\% of signal events.

\section{Event selection}

Signal events are characterized by a high value of \ptmiss, the absence of any isolated lepton (e, $\mu$, or $\tau$) or photon, and the presence of a CA15 jet identified as a Higgs boson candidate.
In the signal region (SR) described below, the dominant background contributions arise from {\cPZ}+jets, {\PW}+jets, and \ttbar production.
To predict the \ptmiss~spectra of these processes in the SR, data from different control regions (CRs) are used.
Single-lepton CRs are designed to predict the \ttbar and {\PW}+jets backgrounds, while dilepton CRs predict the {\cPZ}+jets background contribution.
The hadronic recoil, $U$, serves as a proxy for the \ptmiss distribution of the main background processes in the SR and is defined by excluding the electron(s) and muon(s) from the \ptmiss computation in the CRs.
Predictions for other backgrounds are obtained from simulation.

Events are selected online by the high level trigger system, using a jet reconstruction algorithm and constituents that mirror those of the offline analysis. The trigger requires large values of $p_\text{T,trig}^\text{miss}$ or \mht, where $p_\text{T,trig}^\text{miss}$ is the magnitude of the vectorial $\ptvec$ sum over all PF particles and \mht is the magnitude of the vectorial $\ptvec$ sum over all AK4 jets with $\pt>20\GeV$ and $\abs{\eta}<5.2$ at the trigger level.
Muon candidates are excluded from the online $p_\text{T,trig}^\text{miss}$ calculation.
Minimum thresholds on $p_\text{T,trig}^\text{miss}$ and \mht are between 90 and 120\GeV, depending on the data-taking period.
Collectively, online requirements on $p_\text{T,trig}^\text{miss}$ and \mht are referred to as \ptmiss triggers.
These triggers are measured to be 96\% efficient for $\ptmiss (U)>200\GeV$ and 100\% efficient for $\ptmiss(U)>350\GeV$.
For CRs that require the presence of electrons, events are collected by single-electron triggers, in which at least one electron is required by the online selection criteria.
These sets of requirements are referred to as single-electron triggers.

A common set of preselection criteria is used for all regions.
The presence of exactly one CA15 jet with $\pt>200\GeV$ and $\abs{\eta}<2.4$ is required.
It is also required that $100<m_{\mathrm{SD}}<150\GeV$ and $N_2^{\mathrm{DDT}}<0$.
In the SR (CRs), \ptmiss~($U$) has to be larger than 200\GeV, and the minimum azimuthal angle $\phi$ between any AK4 jet and the direction of \ptvecmiss ($\vec{U}$) must be larger than 0.4 radians to reject multijet events that mimic signal events.
Events with any $\tau_\text{had}$ candidate or photon candidate are vetoed.
The number of AK4 jets for which $\Delta R=\sqrt{(\Delta \eta)^2+(\Delta \phi)^2}>1.5$, where $\Delta\eta$ and $\Delta\phi$ are, respectively, the differences in pseudorapidity and in the azimuthal angle (measured in radians) of a given AK4 jet and the CA15 jet, is required to be smaller than two. This number is referred to as ``additional AK4 jets'' in the following.
This requirement significantly reduces the contribution from \ttbar events in the SR.

Events that meet the preselection criteria described above are split into the SR and the different CRs based on their lepton multiplicity and the presence of a {\cPqb}-tagged AK4 jet not overlapping with the CA15 jet, as summarized in Table~\ref{tab:event_selection}.
For the SR, events are selected if they have no isolated electrons (muons) with $\pt >10\GeV$ and $\abs{\eta}< 2.5$ (2.4), and the previously described double-{\cPqb} tag requirement on the Higgs boson candidate CA15 jet is imposed.

\begin{table*}[t]
  \centering
  \topcaption{Event selection criteria defining the signal and control regions. These criteria are applied in addition to the preselection common to all regions,
as described in the text. The presence of a \cPqb-tagged AK4 jet that does not overlap with the CA15 jet is vetoed in all analysis regions except for the single-lepton CR enriched in \ttbar events, for which such an AK4 \cPqb~tag is required.} \label{tab:event_selection}
  \cmsTable{
    \begin{tabular}{  l  c  c  c  c }
      \hline
        Region   & Main background process & Additional AK4 {\cPqb} tag   & Leptons & Double-{\cPqb} tag \\ \hline
        Signal   & {\cPZ}+jets, \ttbar, {\PW}+jets & 0                & 0       & pass \\
        Single-lepton        & {\PW}+jets, \ttbar & 0                & 1       & pass/fail\\
        Single-lepton, {\cPqb}-tagged   &  \ttbar, {\PW}+jets & 1                & 1       & pass/fail\\
        Dilepton & {\cPZ}+jets & 0                & 2       & pass/fail\\
      \hline
    \end{tabular}
  }
\end{table*}

To predict the \ptmiss spectrum of the {\cPZ}+jets process in the SR, dimuon and dielectron CRs are used.
Dimuon events are selected online employing the same \ptmiss triggers that are used in the SR.
These events are required to have two oppositely charged muons (having $\pt >20\GeV$ and $\pt > 10\GeV$ for the leading and trailing muon, respectively) with an invariant mass between 60 and 120\GeV.
The leading muon has to satisfy tight identification and isolation requirements and is selected with an average efficiency of 95\%.
Dielectron events are selected online using single-electron triggers.
Two oppositely charged electrons with \pt greater than 10\GeV are required offline, and they must form an invariant mass between 60 and 120\GeV.
To be on the plateau of the trigger efficiency, at least one of the two electrons must have $\pt>40\GeV$  and must satisfy tight identification and isolation requirements that correspond to an efficiency of 70\%~\cite{Khachatryan:2015hwa}.

Events that satisfy the SR selection because of the loss of a single lepton primarily originate from {\PW}+jets and semileptonic \ttbar events.
To predict these backgrounds, four single-lepton samples are used: single-electron and single-muon, with and without a {\cPqb}-tagged AK4 jet outside the CA15 jet.
The single-lepton CRs with a {\cPqb}-tagged AK4 jet target \ttbar events, while the other two single-lepton CRs target {\PW}+jets events.
Single-muon events are selected using the \ptmiss triggers described above. Single-electron events are selected using the same single-electron triggers employed in the online selection of dielectron events.
The electron (muon) candidate in these events is required to have $\pt > 40$ (20)\GeV and to satisfy tight identification and isolation requirements.
In addition, samples with a single electron must have $\ptmiss>50\GeV$ to avoid a large contamination from multijet events.

Each CR is further split into two subsamples depending on whether or not the CA15 jet satisfies the double-{\cPqb} tag requirement.
This division allows for an in situ calibration of the scale factor that corrects the simulated misidentification probability of the double-{\cPqb} tagger for the three main backgrounds to the probability observed in data.

\section{Signal extraction}

As mentioned in Section~\ref{intro}, signal and background contributions to the data are extracted with a simultaneous binned likelihood fit (using the \textsc{RooStats} package~\cite{RooStats}) to the \ptmiss and $U$ distributions in the SR and the CRs.
The dominant SM process in each CR is used to predict the respective background in the SR via transfer factors $T$.
These factors are determined in simulation and are given by the ratio of the prediction for a given bin in \ptmiss in the SR and the corresponding bin in $U$ in the CR, for the given process.
This ratio is determined independently for each bin of the corresponding distribution.

For example, if b$\ell$ denotes the \ttbar process in the {\cPqb}-tagged single-lepton control sample that is used to estimate the \ttbar contribution in the SR, the expected number of \ttbar events, $N_{i}$, in the $i^\text{th}$ bin of the SR is then given by $N_{i}= \mu^{\ttbar}_{i}/T^{\mathrm{b}\ell}_{i}$, where  $\mu^{\ttbar}_i$ is a freely floating parameter included in the likelihood to scale the \ttbar contribution in bin $i$ of $U$ in the CR.

The transfer factors used to predict the {\PW}+jets and \ttbar backgrounds take into account the impact of lepton acceptances and efficiencies, the {\cPqb} tagging efficiency, and, for the single-electron control samples, the additional requirement on \ptmiss.
Since the CRs with no {\cPqb}-tagged AK4 jets and a double-{\cPqb}-tagged CA15 jet also have significant contributions from the \ttbar process,  transfer factors to predict this contamination from \ttbar events are also imposed between the single-lepton CRs with and without {\cPqb}-tagged AK4 jets.
A similar approach is applied to estimate the contamination from {\PW}+jets production in the \ttbar CR with events that fail the double-{\cPqb} tag requirement.
Likewise, the {\cPZ}+jets background prediction in the signal region is connected to the dilepton CRs via transfer factors.
They account for the difference in the branching fractions of the ${\cPZ}\to \nu\nu$~and the ${\cPZ}\to \ell\ell$ decays and the impacts of lepton acceptances and selection efficiencies.

\section{Systematic uncertainties}

Nuisance parameters are introduced into the likelihood fit to represent the systematic uncertainties of the search.
They can affect either the normalization or the shape of the \ptmiss ($U$) distribution for a given process in the SR (CRs) and can be constrained in the fit.
The shape uncertainties are incorporated by means of Gaussian prior distributions, while the rate uncertainties are given a log-normal prior distributions.
The list of the systematic uncertainties considered in this search is presented in Table~\ref{tab:systs}.
To better estimate their impact on the results, uncertainties from a similar source (\eg, uncertainties in the trigger efficiencies) have been grouped.
The groups of uncertainties have been ordered by the improvement in sensitivity obtained by removing the corresponding nuisances in the likelihood fit.
The sensitivity in the baryonic \cPZpr\ model is generally poorer than that of a 2HDM+\Pa model because the former predicts a more background-like \ptmiss~distribution.
The description of each single uncertainty in the text follows the same order as in the table.

Scale factors are used to correct for differences in the double-{\cPqb} tagger misidentification efficiencies in data and in the simulated {\PW}/{\cPZ}+jets and \ttbar samples.
These scale factors are measured by simultaneously fitting events that pass or fail the double-{\cPqb} tag requirement.
The correlation between the double-{\cPqb} tagger and \ptmiss (or $U$) is taken into account by allowing recoil bins to fluctuate within a constraint that depends on the recoil value.
Such dependence is estimated from the profile of the two-dimensional distribution of the double-{\cPqb} tag discriminant vs. the $\pt$ of the CA15 jet.
This is the shape uncertainty that has the largest impact on the upper limits on the signal cross sections.

Shape uncertainties due to the bin-by-bin statistical uncertainties in the transfer factors are considered for the {\cPZ}+jets, {\PW}+jets, and \ttbar processes.

For the signal and the SM \Ph processes, an uncertainty in the double-{\cPqb} tagging efficiency is applied that depends on the \pt of the CA15 jet.
This shape uncertainty has been derived through a measurement performed using a sample enriched in multijet events with double-muon-tagged $\Pg\to\bbbar$ splittings.
A 7\% rate uncertainty in the efficiency of the requirement on the substructure variable $N_2^\mathrm{DDT}$, which is used to identify two-prong CA15 jets, is assigned to all processes where the decay of a resonance inside the CA15 jet cone is expected.
Such processes include signal production together with SM \Ph and diboson production.
The uncertainty has been derived from the efficiency measurement obtained by performing a fit in a control sample enriched in semi-leptonic \ttbar events, where the CA15 jet originates from the {\PW} boson that comes from the hadronically decaying top quark.

A 4\% rate uncertainty due to the imperfect knowledge of the CA15 jet energy scale~\cite{jec} is assigned to all the processes obtained from simulation.

Similarly, a 5\% rate uncertainty in the \ptmiss magnitude, as measured by CMS in Ref.~\cite{Khachatryan:2014gga}, is assigned to each of the processes estimated from simulation.

A rate uncertainty of 2.5\% in the integrated luminosity measurement~\cite{CMS-PAS-LUM-17-001} is included and assigned to processes determined from simulation.
In these cases, uncertainties in the PDFs and uncertainties due to variations in the QCD renormalization and factorization scales are included as shape uncertainties, obtained by varying those parameters in the simulation.

\begin{table*}[t]
  \centering
    \topcaption{Sources of systematic uncertainty, along with the type (rate/shape)
      of uncertainty and the affected processes. For the rate uncertainties,
      the percentage value of the prior is quoted. The last column denotes the improvement in the expected limit when
      removing the uncertainty group from the list of nuisances included
      in the likelihood fit. Such improvement is estimated considering as signal processes the 2HDM+\Pa model with $m_\PSA=1.1\TeV$ and $m_\Pa=150\GeV$ and the baryonic {\cPZpr} model with $m_{\cPZpr}=0.2\TeV$ and $m_\chi=50\GeV$.}
   \cmsTable{
    \begin{tabular}{l c c c c}
      \hline
      Systematic uncertainty & Type & Processes & \multicolumn{2}{c}{Impact on sensitivity}  \\
       &  &  &  2HDM+\Pa & Baryonic \cPZpr  \\
      \hline
      Double-{\cPqb} mistagging & shape & {\cPZ}+jets, {\PW}+jets, \ttbar & 4.8\% & 14.8\%\\[\cmsTabSkip]
      Transfer factor stat. uncertainties & shape & {\cPZ}+jets, {\PW}+jets, \ttbar & 1.9\%  & 4.0\%\\[\cmsTabSkip]
      Double-{\cPqb} tagging & shape & SM \Ph, signal & \multirow{ 2}{*}{1.2\%} & \multirow{ 2}{*}{1.1\%}\\
      $N_2^\mathrm{DDT}$ efficiency & 7\% & diboson, SM \Ph, signal \\[\cmsTabSkip]
      CA15 jet energy & 4\% & single {\cPqt}, diboson, multijet, SM \Ph, signal  & 0.8\% & 0.6\%\\[\cmsTabSkip]
      \ptmiss magnitude & 5\% & all & 0.7\% & $<$0.5\%\\[\cmsTabSkip]
      Integrated luminosity & 2.5\% & single {\cPqt}, diboson, multijet, SM \Ph, signal &$<$0.5\% & $<$0.5\%\\[\cmsTabSkip]
      $\ptmiss$ trigger efficiency & shape/rate & all & \multirow{ 2}{*}{$<$0.5\%} & \multirow{ 2}{*}{$<$0.5\%} \\
      Single-electron trigger & 1\% & all \\[\cmsTabSkip]
      AK4 {\cPqb} tagging & shape & all & $<$0.5\% & $<$0.5\% \\[\cmsTabSkip]
      $\tau$ lepton veto & 3\% & all &\multirow{2}{*}{$<$0.5\%} & \multirow{2}{*}{$0.7\%$}\\
      Lepton efficiency & 1\% per lepton & all \\[\cmsTabSkip]
      Renorm./fact. scales & shape & SM \Ph &\multirow{4}{*}{$<$0.5\%} & \multirow{4}{*}{$<$0.5\%}\\
      PDF & shape & SM \Ph \\
      Multijet normalization & 100\% & multijet \\
      Theoretical cross section & 20\% & single {\cPqt}, diboson\\
      \hline
    \end{tabular}
   }
   \label{tab:systs}
\end{table*}

The \ptmiss trigger efficiency is parametrized as a function of $U$ and measured using both single-muon and dimuon events.
The difference between these measurements is used to derive an uncertainty, which results in a 1\% rate uncertainty for processes estimated using simulation.
Processes estimated using control regions (\ttbar, {\PW}+jets, and {\cPZ}+jets) are sensitive to the effect of this uncertainty as a function of $U$, so a shape uncertainty (as large as 2\% at low $U$ values) is considered for such processes.
The efficiencies of the single-electron triggers are parametrized as a function of the electron \pt and $\eta$ and an associated 1\% systematic uncertainty is added into the fit.

An uncertainty in the efficiency of the CSV {\cPqb} tagging algorithm applied to isolated AK4 jets is assigned to the transfer factors used to predict the \ttbar background.
The scale factors that correct this efficiency are measured with standard CMS methods~\cite{Sirunyan:2017ezt}.
They depend on the \pt and $\eta$ of the {\cPqb}-tagged (or mistagged) jet and therefore their uncertainties are included in the fit as shape uncertainties.

The uncertainty in the $\tau$ lepton veto amounts to 3\%, correlated across all $U$ bins.
Also correlated across all $U$ bins are the uncertainties in the electron and muon selection efficiencies, which amount to 1\%.

An uncertainty of 21\% in the heavy-flavor fraction of {\PW}+jets is reported in previous CMS measurements~\cite{Khachatryan:2014uva,Chatrchyan:2013uza}.
The uncertainty in the heavy-flavor fraction of jets produced together with a {\cPZ} boson is measured to be 22\%~\cite{Khachatryan:2014zya,Chatrchyan:2014dha}.
To take into account the variation of the double-{\cPqb} tagging efficiency introduced by such uncertainties, the efficiencies for the {\PW}+jets and {\cPZ}+jets processes are reevaluated after varying the heavy-flavor component in the simulation.
The difference in the efficiency with respect to the nominal efficiency value is taken as a systematic uncertainty, and amounts to 4\% in the rate of the {\PW}+jets process and 5\% in the rate of the {\cPZ}+jets process.

Uncertainties in the SM \Ph production due to variations of the renormalization/factorization scales and PDFs are included as shape variations.
An uncertainty of 100\% is assigned to the QCD multijet yield.
This uncertainty is estimated using a sample enriched in multijet events.
The sample is obtained by vetoing leptons and photons, requiring $\ptmiss>250\GeV$ and requiring that the minimum azimuthal angle between \ptvecmiss and the jet directions be less than 0.1\,radians.
One nuisance parameter represents the uncertainty in QCD multijet yields in the signal region, while separate nuisance parameters are introduced for the muon CRs and electron CRs.
A systematic uncertainty of 20\% is assigned to the single top quark background yields as reported by CMS in Ref.~\cite{Chatrchyan:1642680} and is correlated between the SR and the CRs.
An uncertainty of 20\%, correlated across the SR and CRs, is also assigned to the diboson production cross section as measured by CMS in Refs.~\cite{Khachatryan:2016txa,Khachatryan:2016tgp}.

\section{Results}

The expected yields for each background in the SR and their uncertainties, as determined in the likelihood fit under the background-only assumption, are presented in Table~\ref{tab:eventYieldTable}, along with the observed data yields.
Good agreement is observed between data and the predictions. Due to anticorrelations between background processes, in some bins the uncertainty in the background sum is smaller than the uncertainties in the individual contributions, such as, for example, the {\cPZ}+jets yields.
\begin{table*}[t!]
\centering
  \topcaption{Post-fit event yield expectations per \ptmiss bin for the SM backgrounds in the signal region when including the signal region data in the likelihood fit, under the background-only assumption. Also quoted are the expected yields for two signal models. Uncertainties quoted in the predictions include both the systematic and statistical components.}
\cmsTable{
\begin{tabular}{l r r r r}
  \hline
\ptmiss bin         & 200--270\GeV          & 270--350\GeV          & 350--475\GeV          & $>$475\GeV         \\
\hline
{\cPZ}+jets          &$ 249\pm22 $       & $97.2\pm8.5$         & $32.6\pm3.6$          & $11.1\pm1.9$       \\
\ttbar          &$ 199\pm14 $       & $52.1\pm5.2$          & $11.1\pm2.0$          & $0.7\pm0.4$        \\
{\PW}+jets          &$ 122\pm22 $       & $45.0\pm8.7$          & $8.4\pm1.9$           & $2.9\pm0.9$            \\
Single {\cPqt}        &$21.0\pm4.2 $          & $6.1\pm1.2$           & $0.9\pm0.2$           & $0.2\pm0.1$         \\
Diboson         &$ 16.0\pm3.1  $        & $7.6\pm1.5$           & $2.4\pm0.5$           & $1.0\pm0.2$ \\
SM \Ph          &$ 12.6\pm1.4 $      & $ 6.6\pm0.7$           & $ 3.3 \pm 0.3$        & $ 1.3\pm 0.1$      \\[\cmsTabSkip]
$\Sigma~(\text{SM})$ & $619\pm20$ & $215 \pm 8$       & $58.7\pm3.7$          & $17.2 \pm 2.0$ \\
Data            & 619       &  214        & 59          &  21 \\ [\cmsTabSkip]
2HDM+\Pa, $m_\PSA=1\TeV$, $m_\Pa=150\GeV$ & $5.7 \pm 0.6$ & $9.8 \pm 1.1$ & $18.5 \pm 2.1$ & $5.2 \pm 0.6$\\
Bar. {\cPZpr}, $m_{\cPZpr}=0.2\TeV$, $m_\chi=50\GeV$ & $184 \pm 20$ & $118 \pm 13$ & $69.5 \pm 7.7$ & $28.9 \pm 3.3$\\
\hline
  \end{tabular}
}
\label{tab:eventYieldTable}
\end{table*}

\begin{figure}[t!]
\centering
 \subfloat{\includegraphics[width=0.475\textwidth]{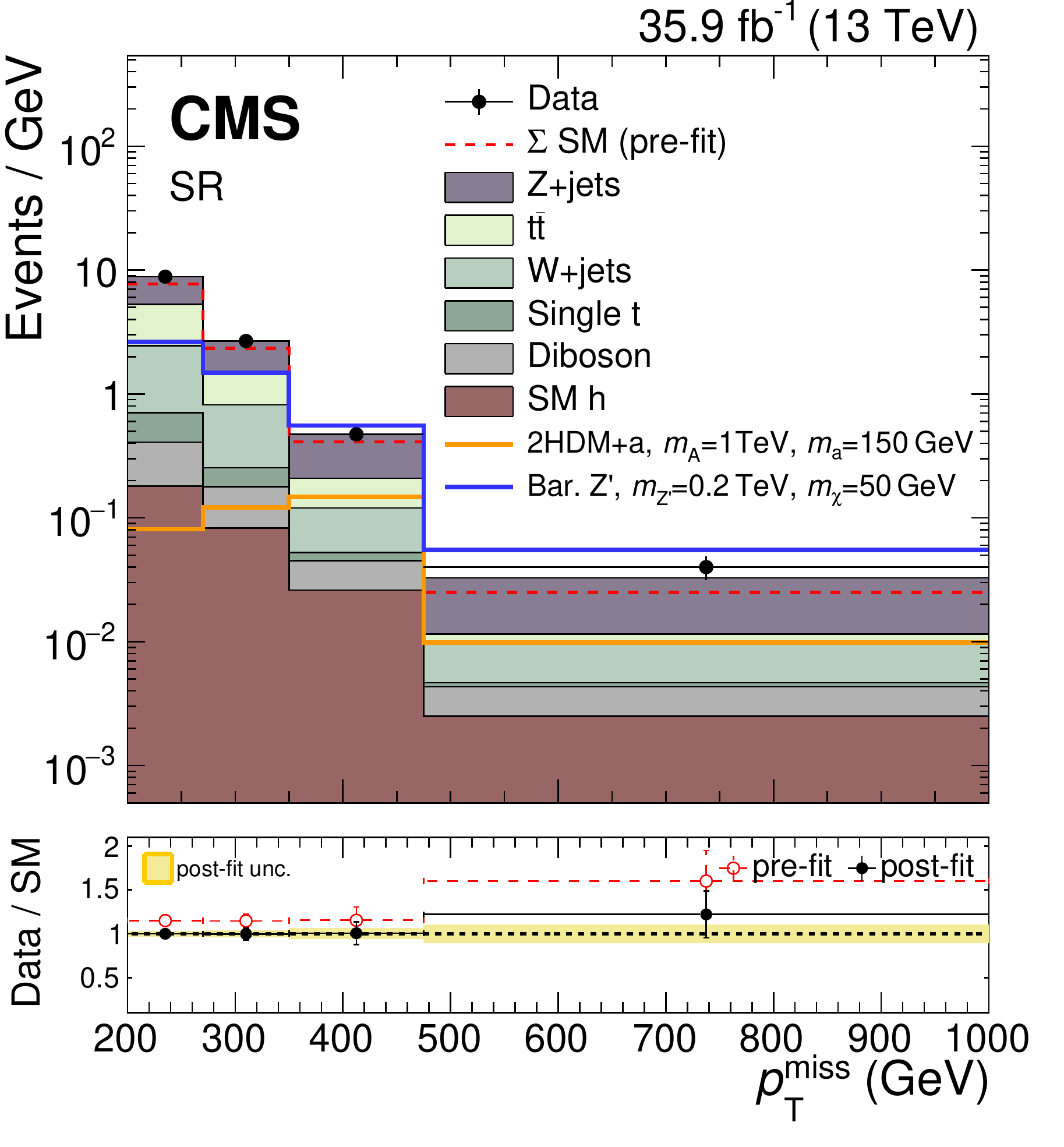}}
\caption{The \ptmiss distribution in the signal region before and after a likelihood fit. The data are in agreement with post-fit background predictions for the SM backgrounds, and no significant excess is observed. The dashed red histogram corresponds to the pre-fit estimate for the SM backgrounds. The lower panel shows the ratio of the data to the predicted SM background, before and after the fit. The rightmost \ptmiss bin includes overflow events.}
\label{Fig_sr}
\end{figure}
Expected yields are also presented for two signal models.
The selection efficiencies for the chosen points correspond to 5\% for the 2HDM+\Pa model and 1\% for the baryonic {\cPZpr} model.

\begin{figure*}
\centering
 \includegraphics[width=0.36\textwidth]{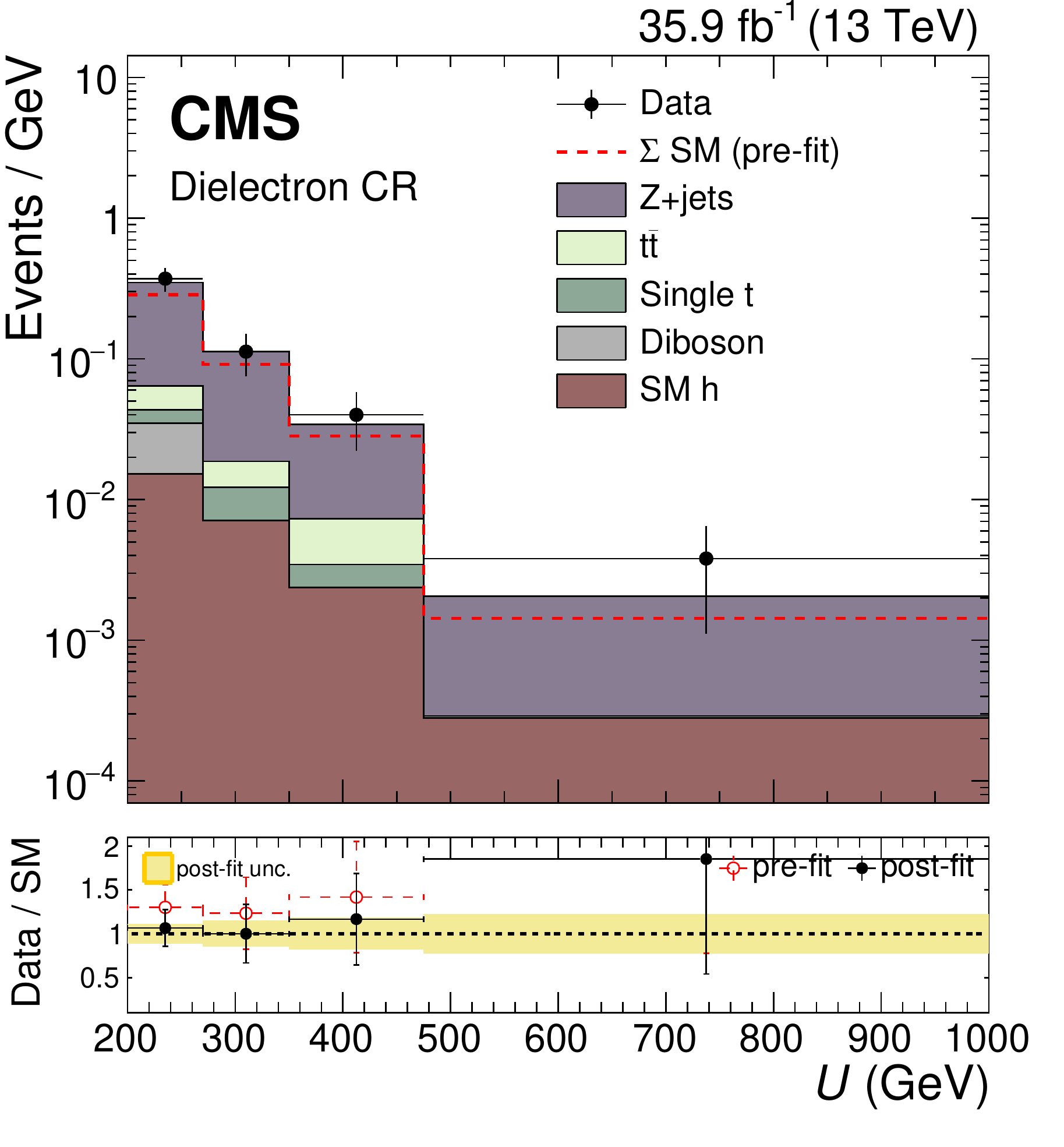}
 \includegraphics[width=0.36\textwidth]{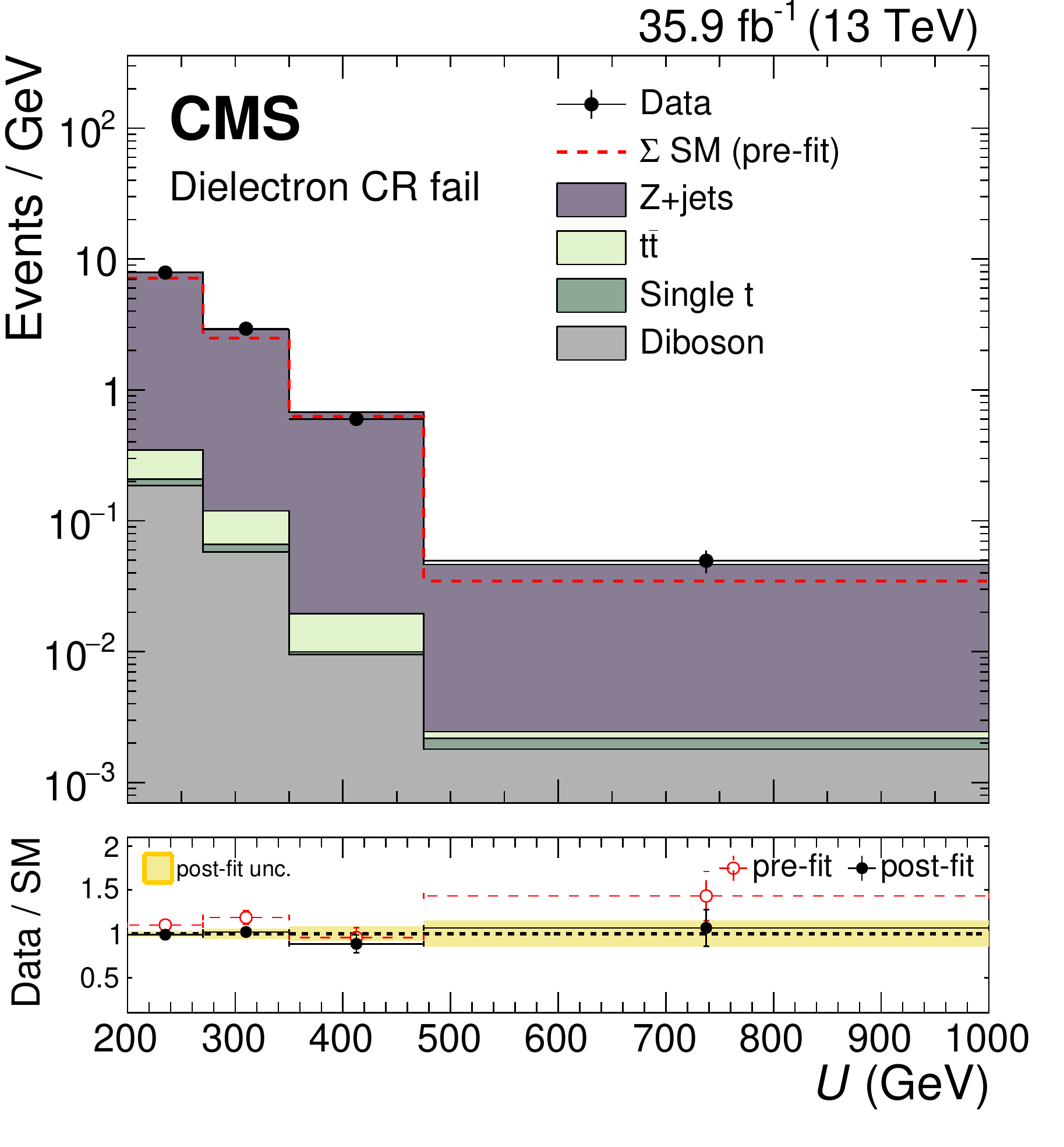} \\
 \includegraphics[width=0.36\textwidth]{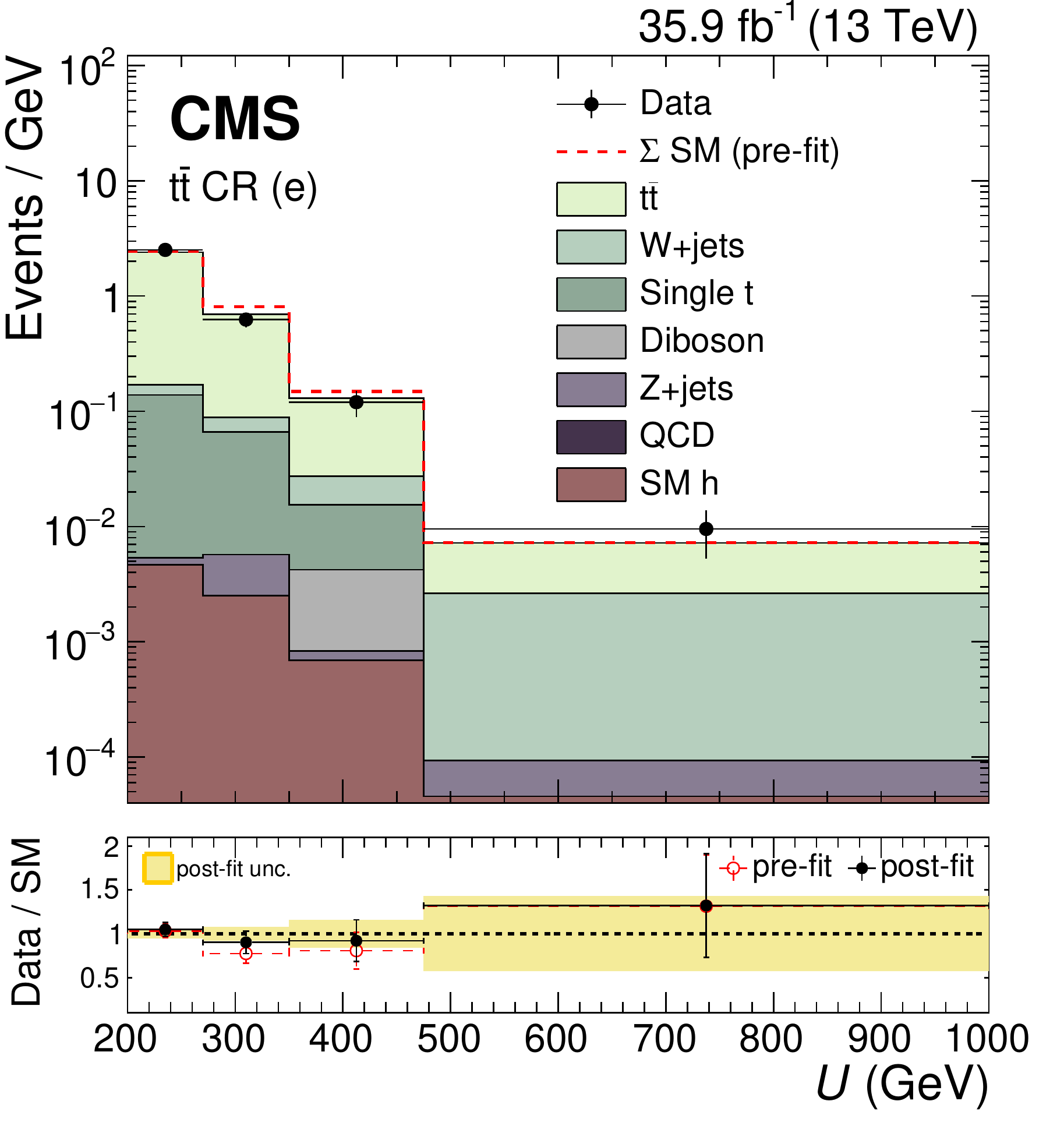}
 \includegraphics[width=0.36\textwidth]{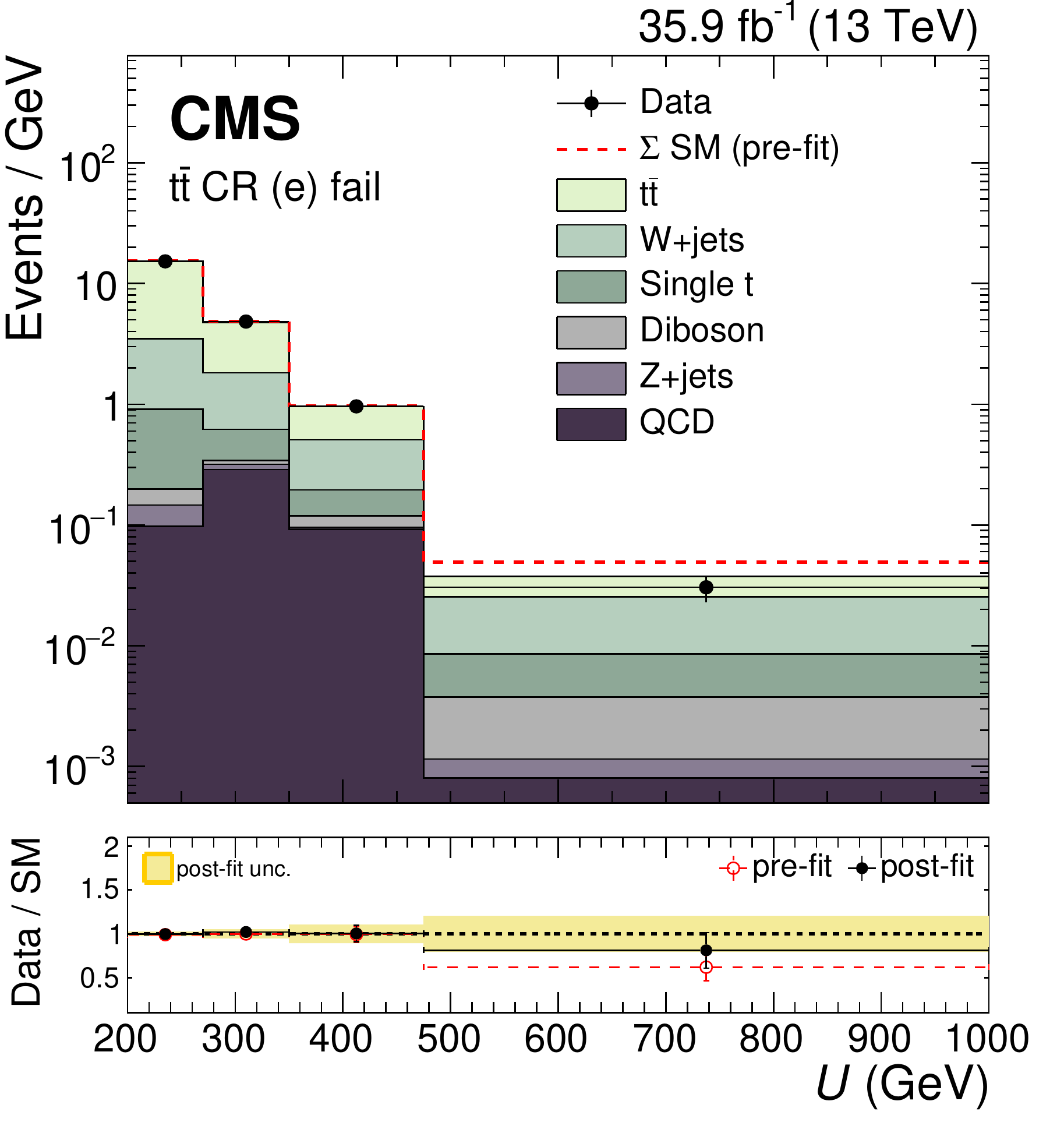} \\
 \includegraphics[width=0.36\textwidth]{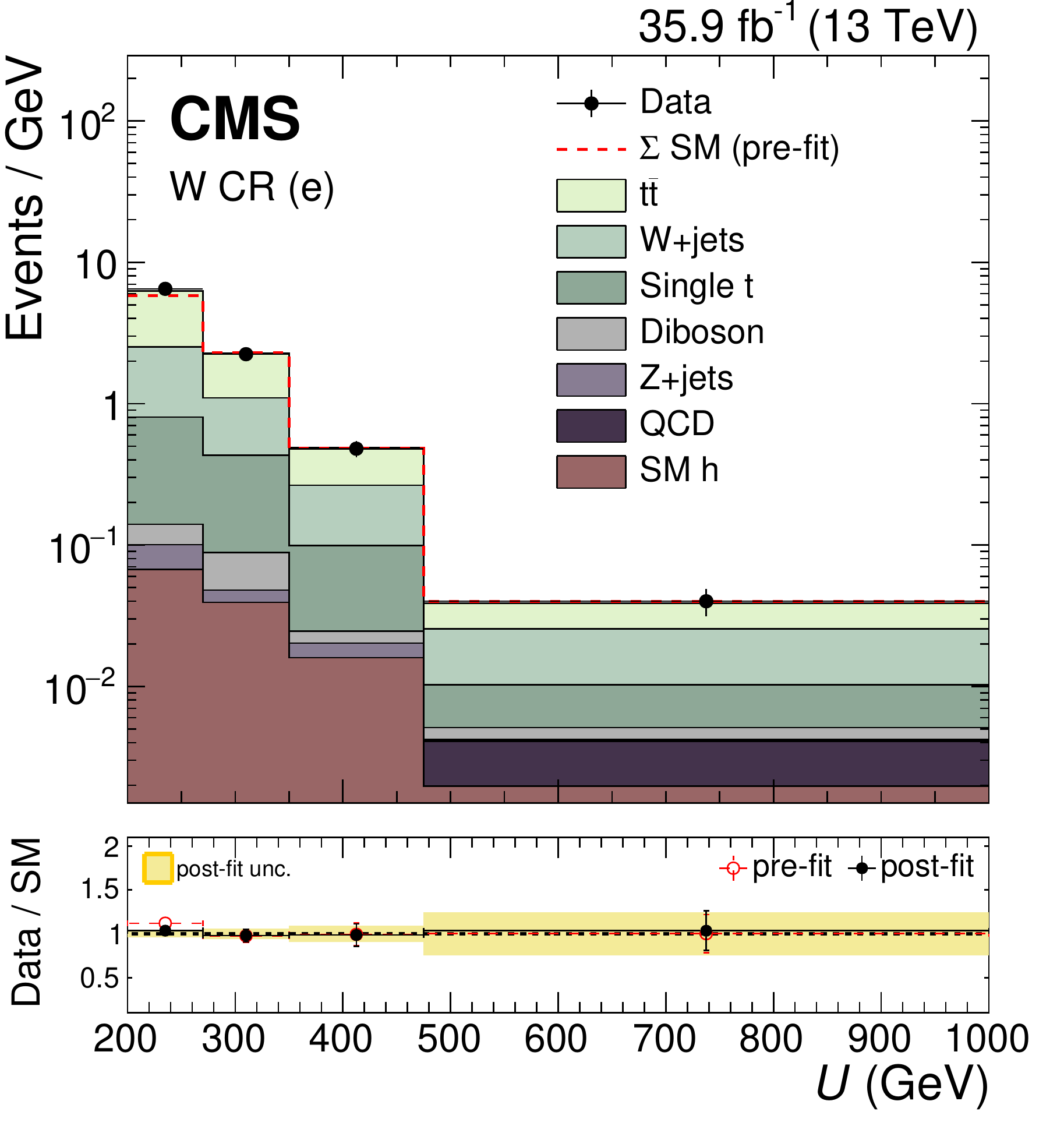}
 \includegraphics[width=0.36\textwidth]{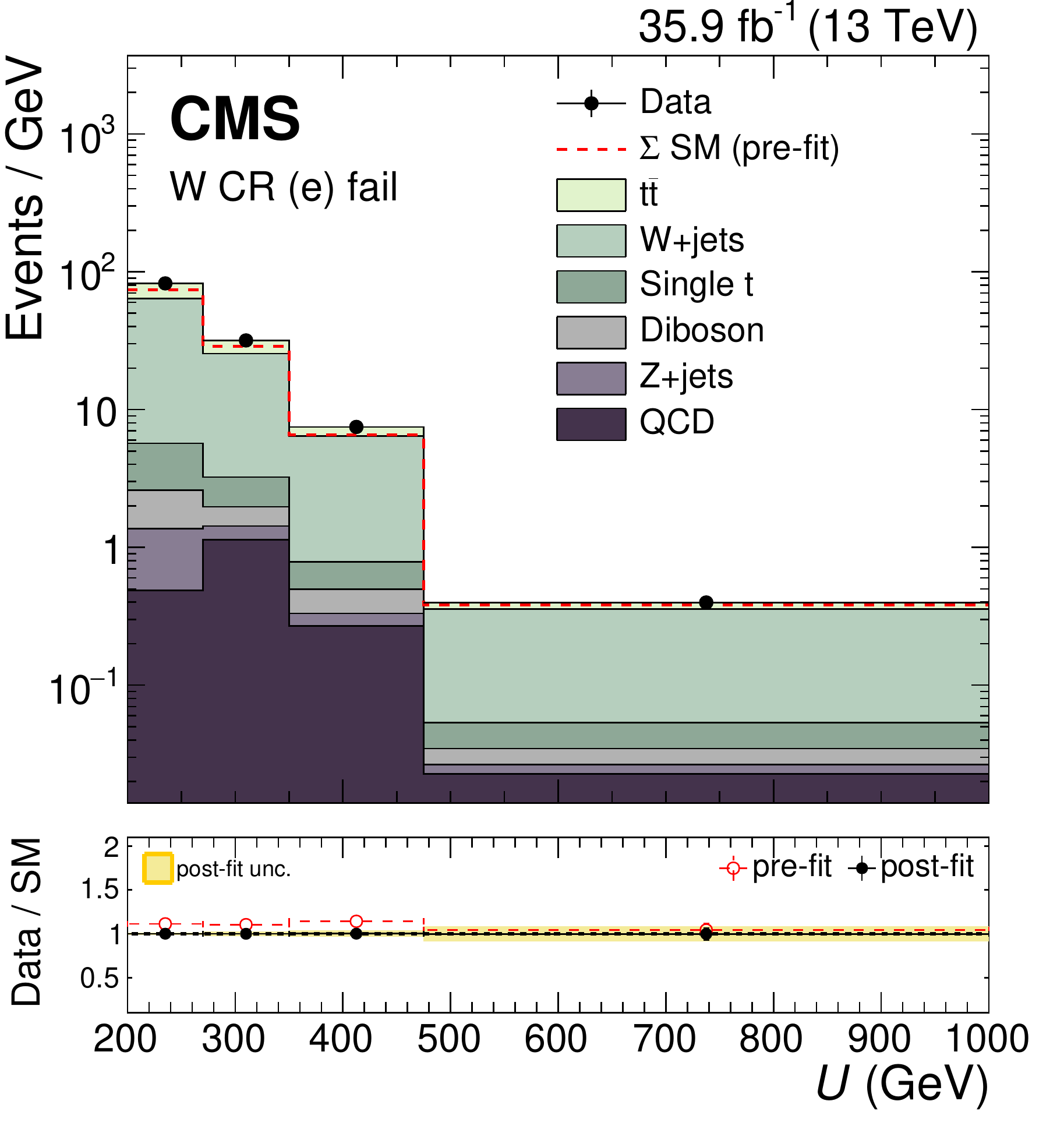}
\caption{The $U$ distribution in the electron control regions before and after a background-only fit to data, including the data in the signal region in the likelihood. For the distributions on the left the CA15 jet passes the double-{\cPqb} tag requirement and for the distributions on the right it fails the double-{\cPqb} tag requirement. The lower panel shows the ratio of the data to the predicted SM background, before and after the fit. The rightmost $U$ bin includes overflow events.}
\label{Fig_cr_1}
\end{figure*}

\begin{figure*}
\centering
 \includegraphics[width=0.36\textwidth]{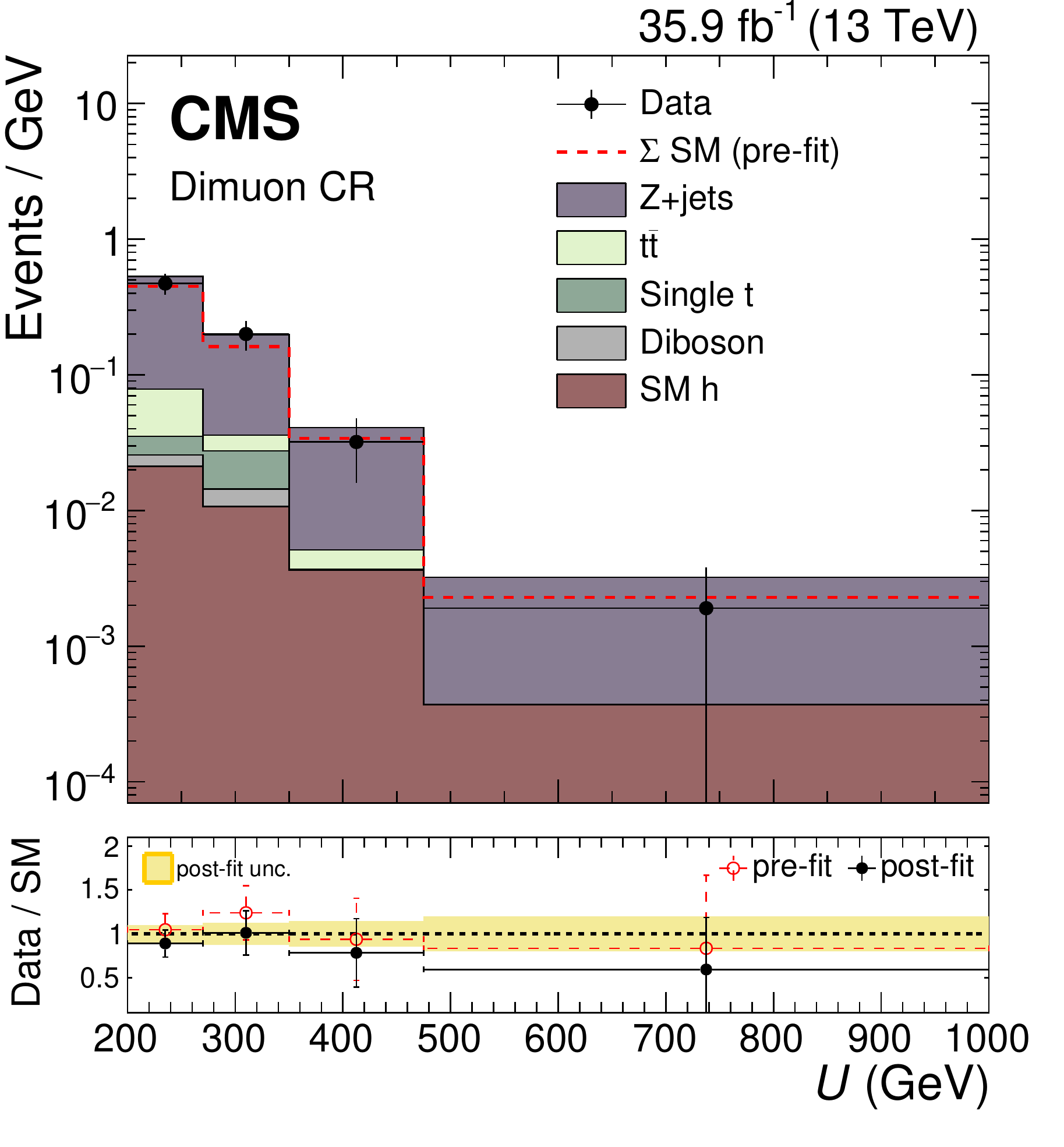}
 \includegraphics[width=0.36\textwidth]{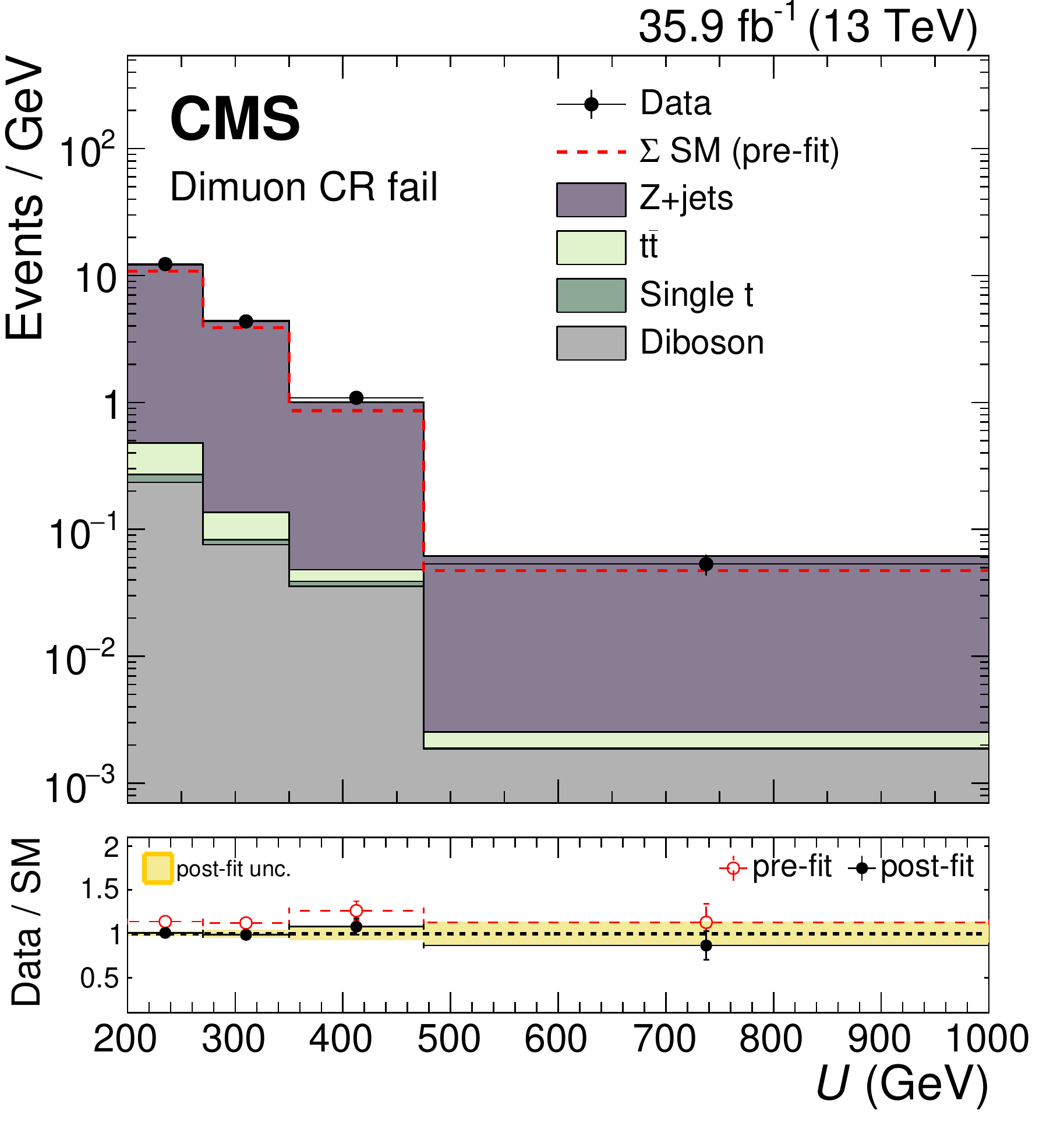} \\
 \includegraphics[width=0.36\textwidth]{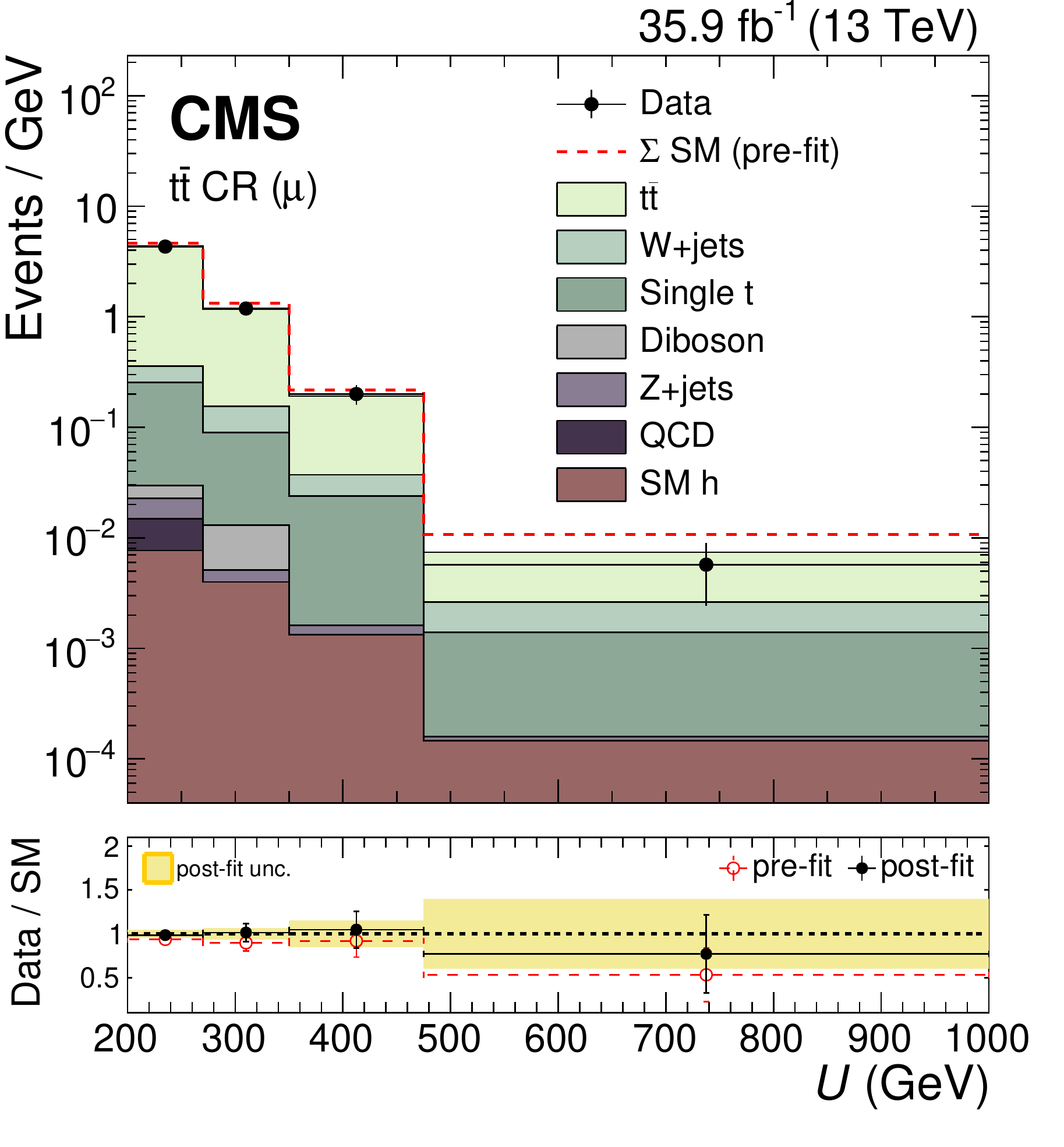}
 \includegraphics[width=0.36\textwidth]{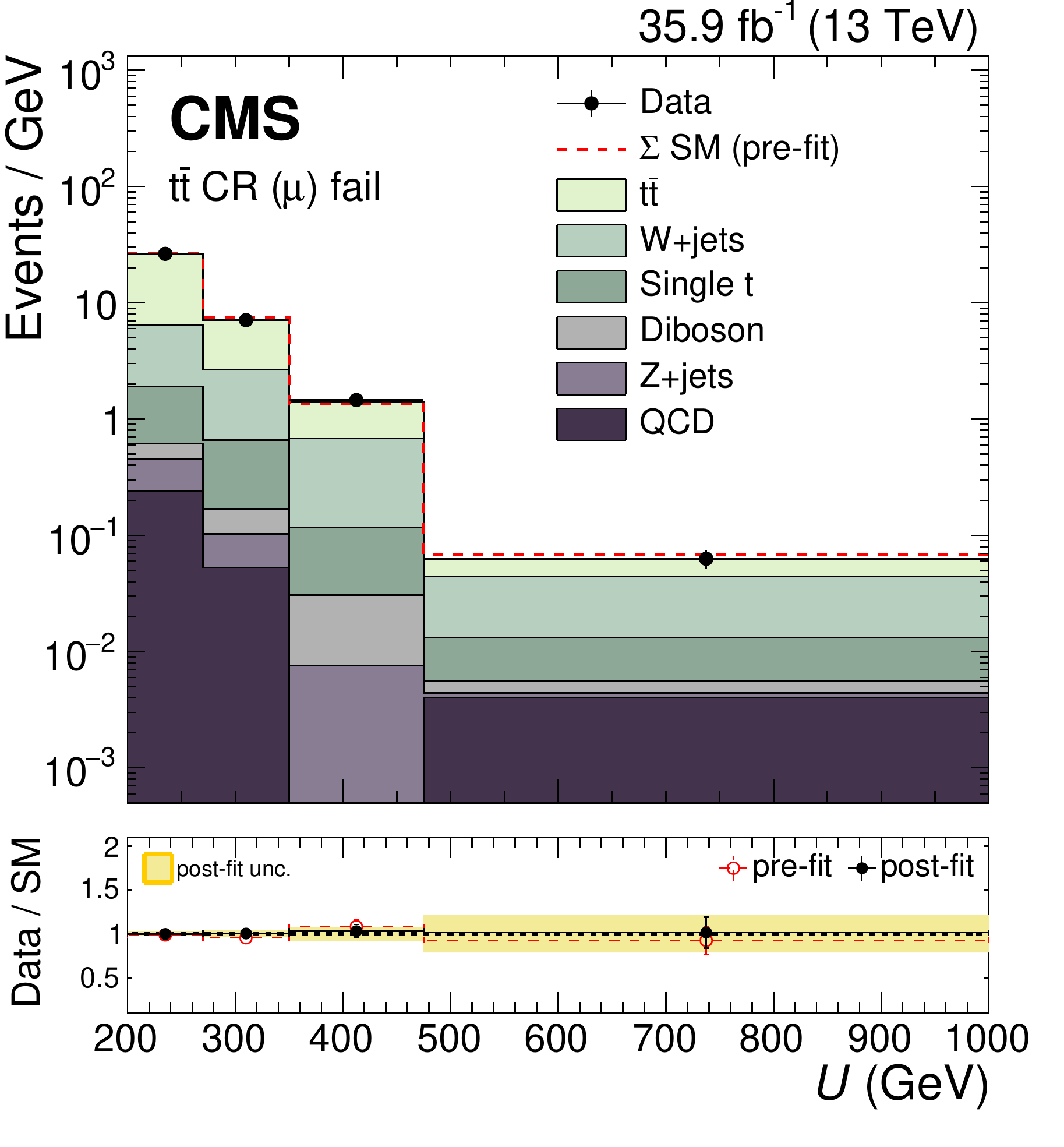}  \\
 \includegraphics[width=0.36\textwidth]{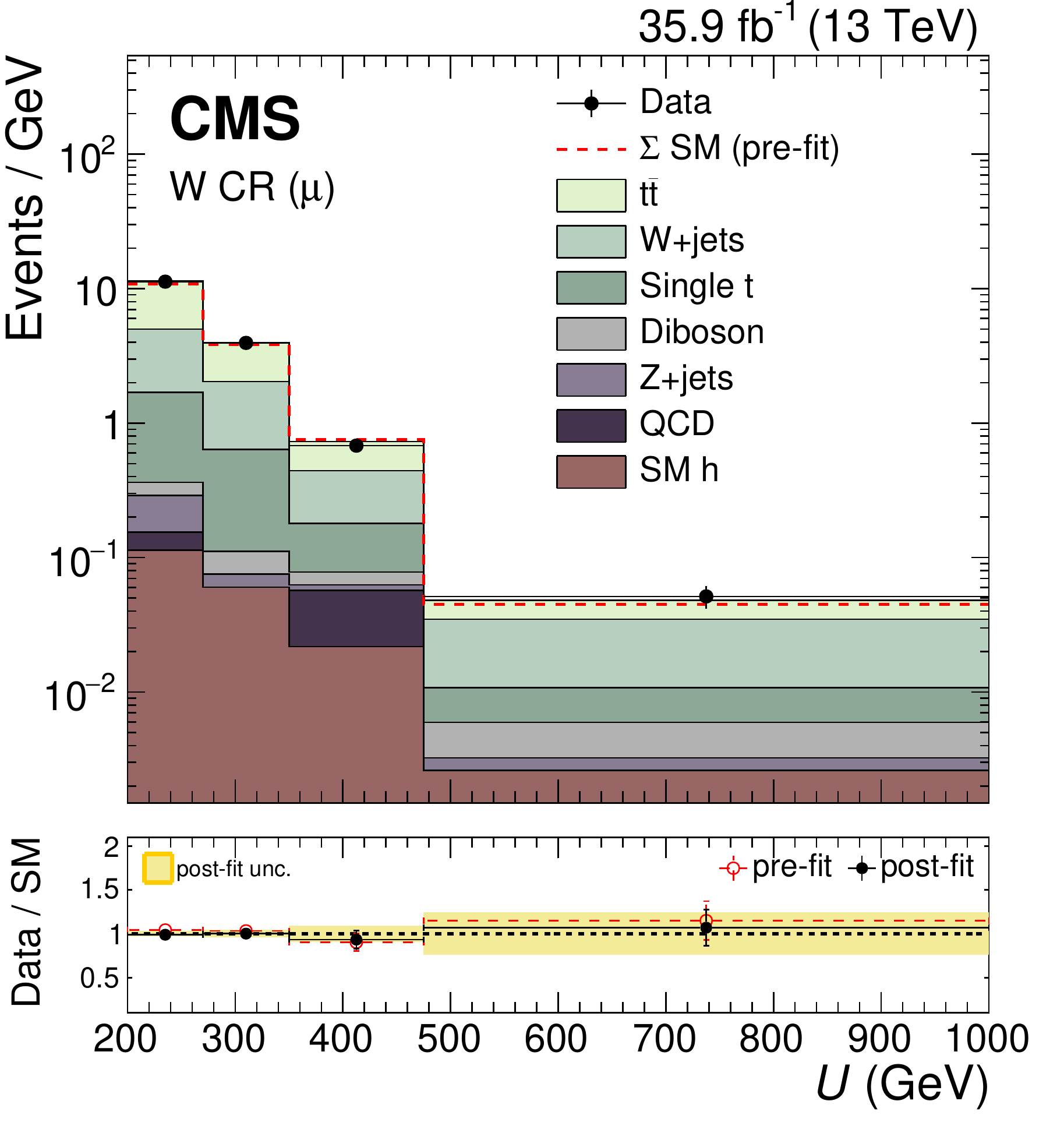}
 \includegraphics[width=0.36\textwidth]{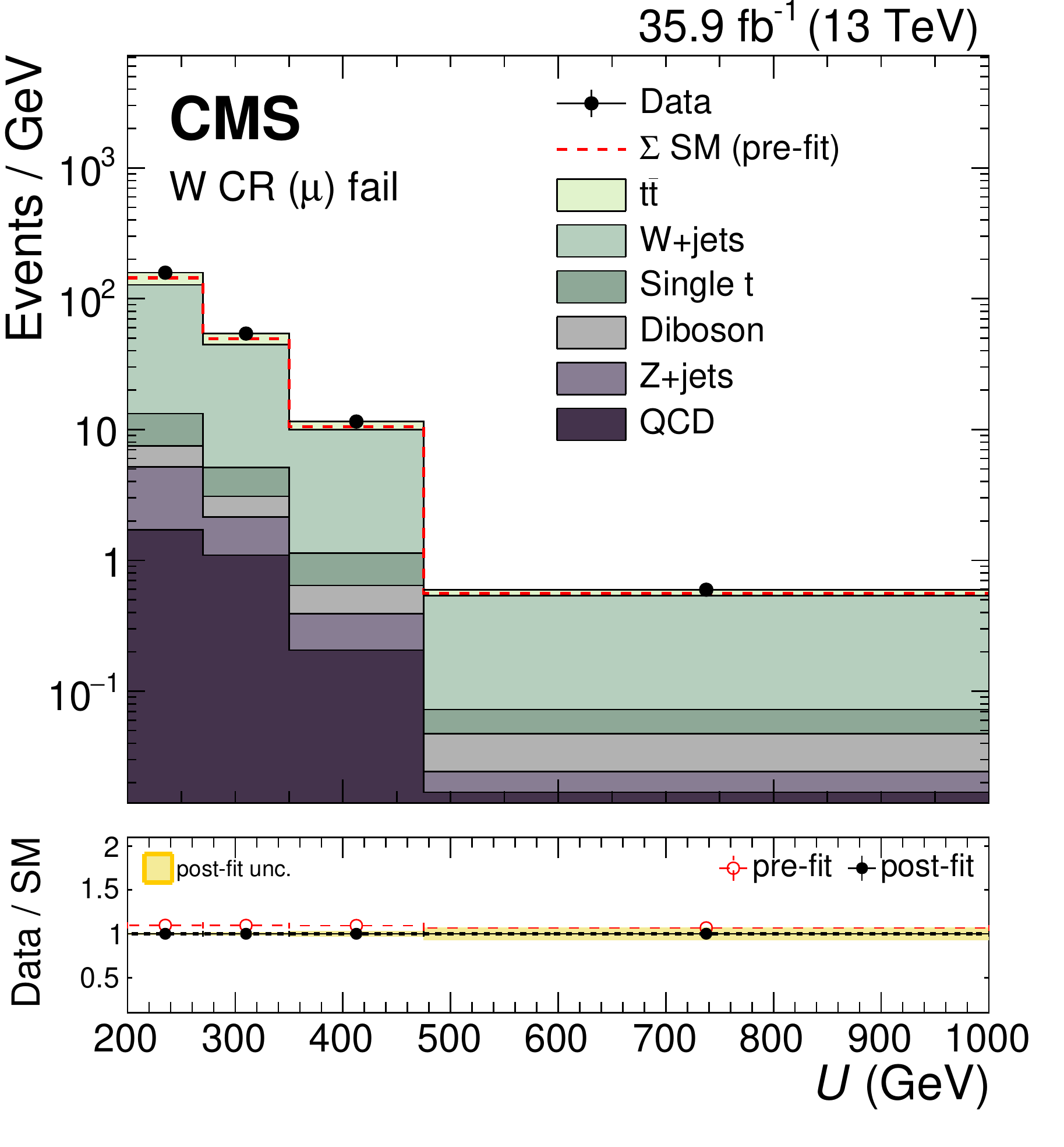}
 \caption{The $U$ distribution in the muon control regions before and after a background-only fit to data, including the data in the signal region in the likelihood. For the distributions on the left the CA15 jet passes the double-{\cPqb} tag requirement and for the distributions on the right it fails the double-{\cPqb} tag requirement. The lower panel shows the ratio of the data to the predicted SM background, before and after the fit. The rightmost $U$ bin includes overflow events.}
\label{Fig_cr_2}
\end{figure*}

Figure \ref{Fig_sr} shows the pre-fit and post-fit \ptmiss distributions in the SR for signal and for all SM backgrounds, as well as the observed data distribution.
The likelihood fit has been performed simultaneously in all analysis regions.
The data agree with the background predictions at the one standard deviation level, and the post-fit estimate of the SM background is slightly larger than the pre-fit one.
The distributions for $U$ in the muon and electron CRs, after a fit to the data, are presented in Figs.~\ref{Fig_cr_1} and \ref{Fig_cr_2}.

No significant excess over the SM background expectation is observed in the SR.
The results of this search are interpreted in terms of upper limits on the signal strength modifier $\mu=\sigma/\sigma_\text{theory}$, where $\sigma_\text{theory}$ is the predicted production cross section of DM candidates in association with a Higgs boson and $\sigma$ is the upper limit on the observed cross section.
The upper limits are calculated at 95\% confidence level (\CL) using a modified frequentist method \cite{Heinemeyer:2013tqa, bib:CLS1, bib:CLS2} computed with an asymptotic approximation \cite{bib:CLS3}.

\begin{figure*}[t]
  \centering
  \includegraphics[width=0.475\textwidth]{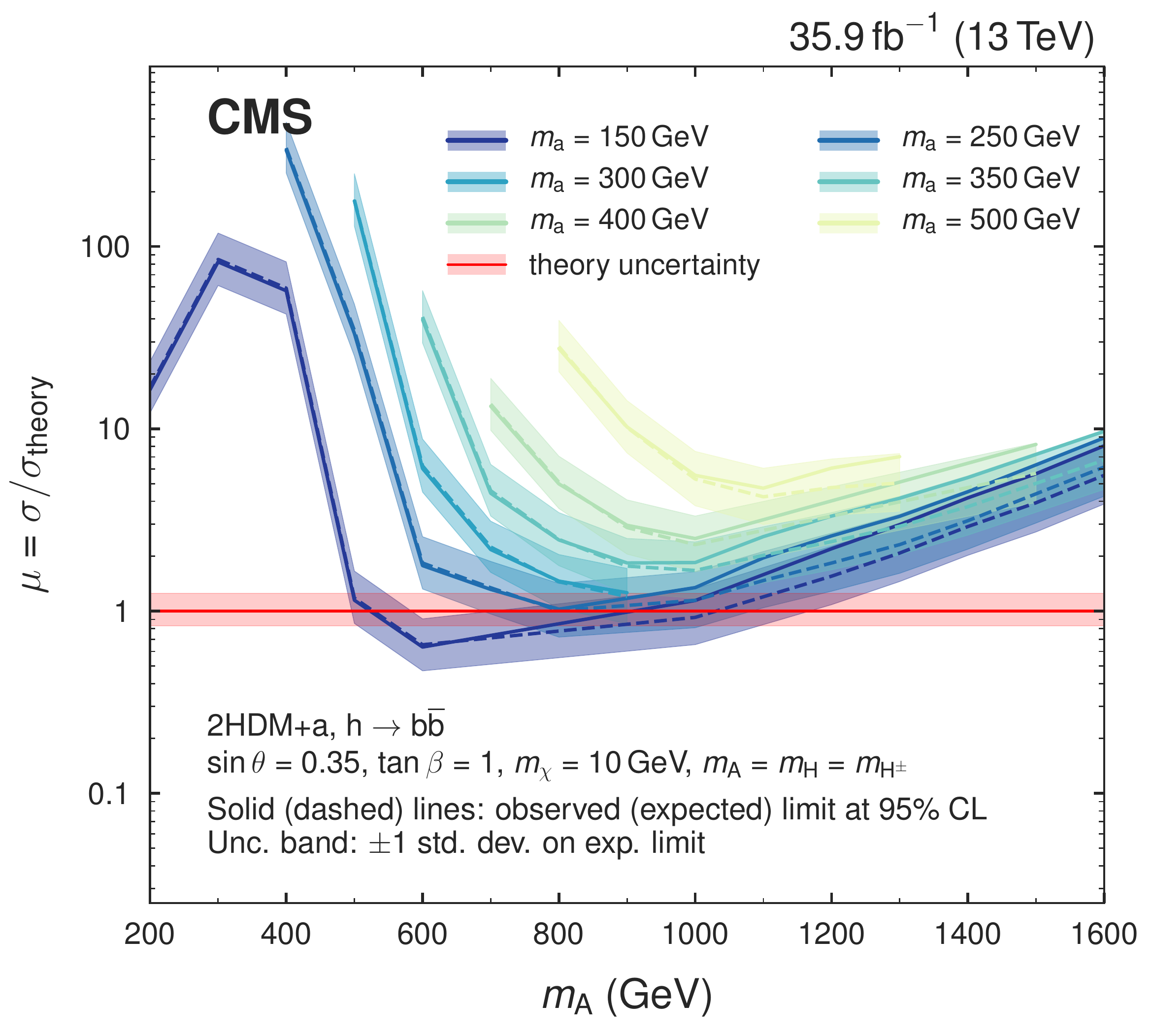}
  \includegraphics[width=0.475\textwidth]{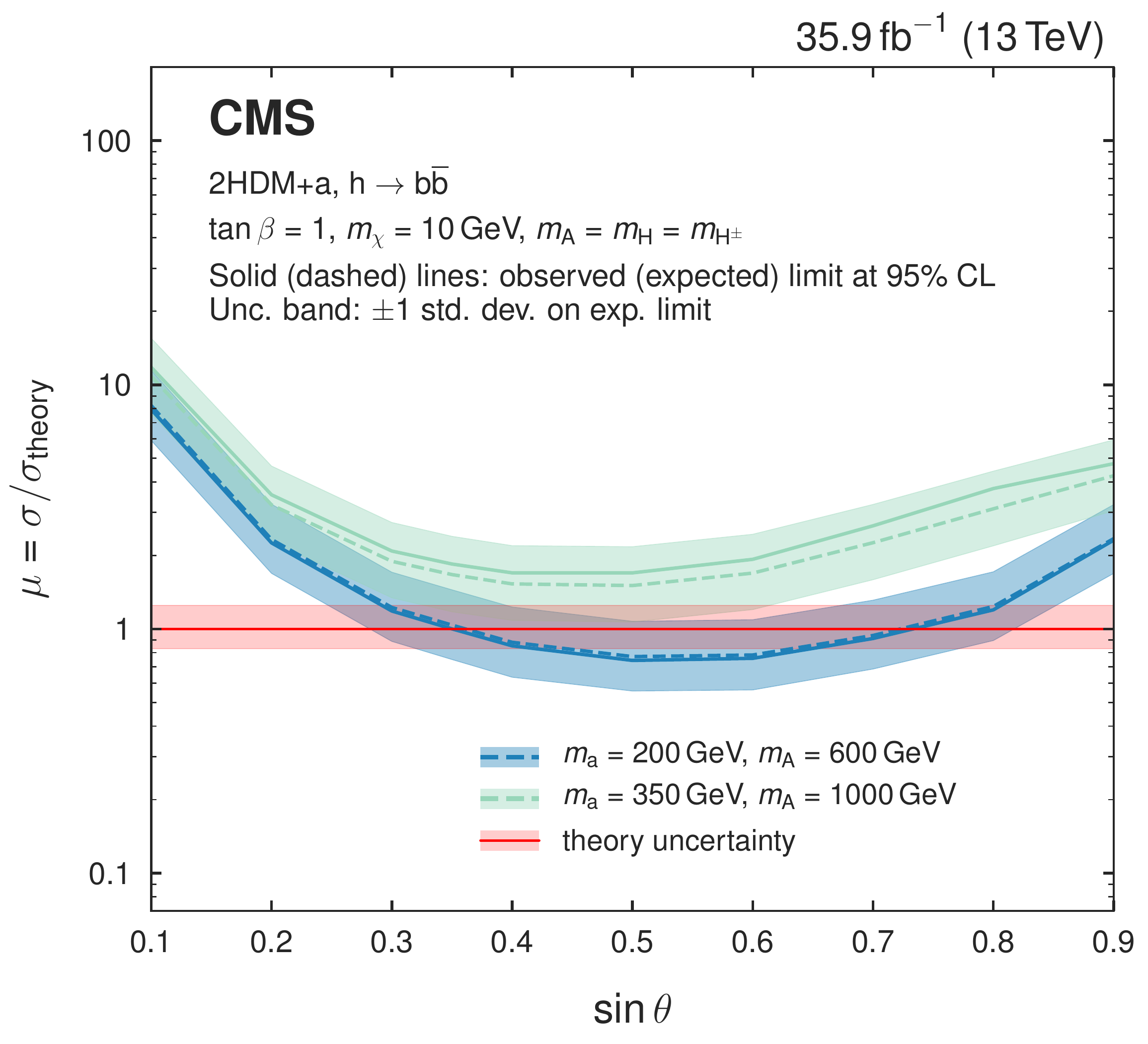}\\
 \includegraphics[width=0.475\textwidth]{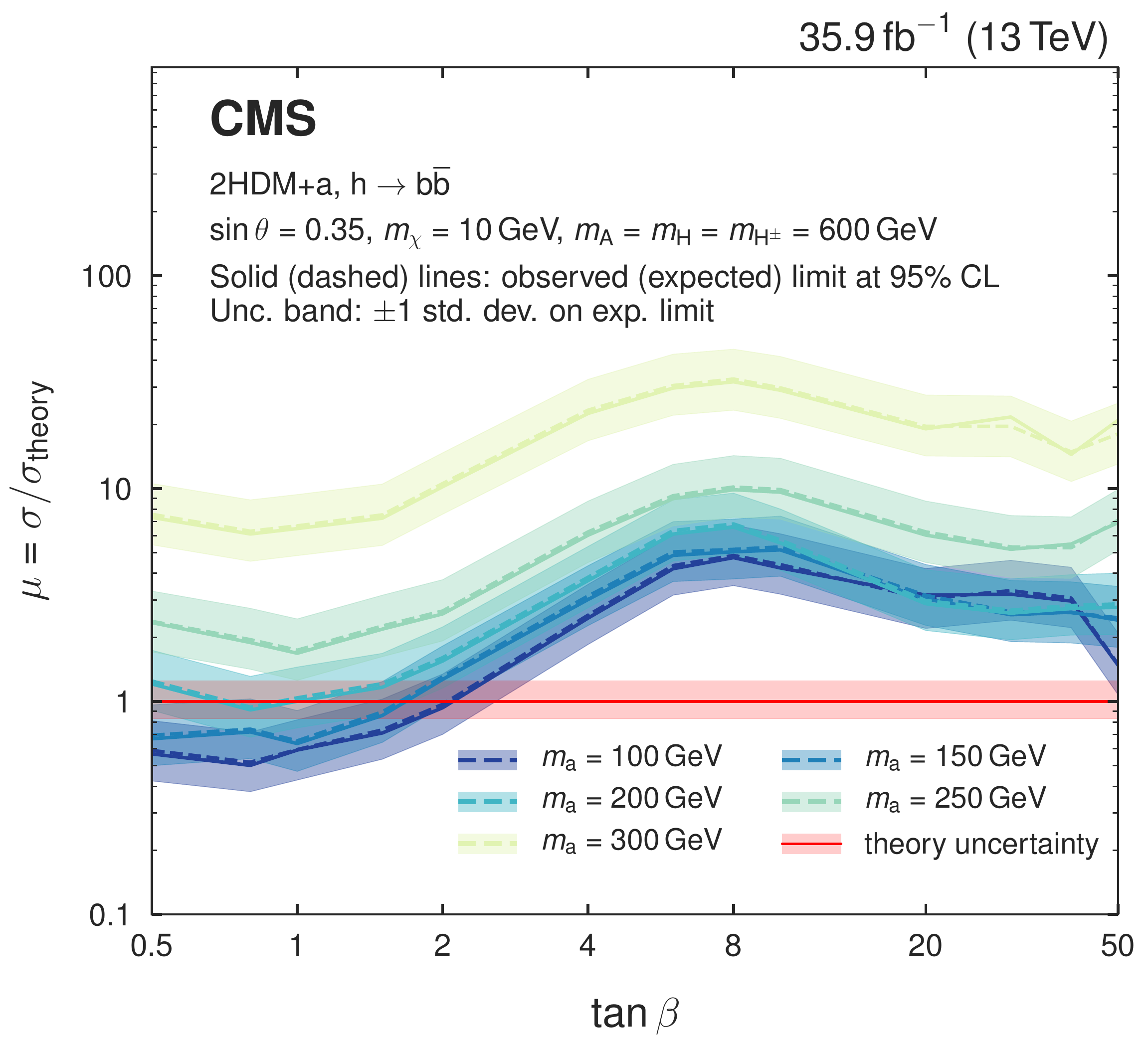}
  \caption{Upper limits at 95\% \CL on the signal strength modifier, defined as $\mu=\sigma/\sigma_\text{theory}$, where $\sigma_\text{theory}$ is the predicted production cross section of DM candidates in association with a Higgs boson and $\sigma$ is the upper limit on the observed cross section. Limits are shown for the 2HDM+\Pa model when scanning $m_\PSA$ and $m_\Pa$ (upper left), the mixing angle $\theta$ (upper right), or $\tan\beta$ (lower). The uncertainty in the computation of $\sigma_\text{theory}$ is 20\% and is shown as a red band around the exclusion line at $\mu=1$.}
  \label{fig:limits_2hdma}
\end{figure*}

Figure~\ref{fig:limits_2hdma} shows the upper limits on $\mu$ for the three scans ($m_\PSA$, $\sin\theta$, and $\tan\beta$) performed.
For the 2HDM+\Pa model, $m_\PSA$ masses are excluded between 500 and 900\GeV for $m_\Pa=150\GeV$, $\sin\theta=0.35$ and $\tan\beta=1$.
Mixing angles with $0.35<\sin\theta<0.75$ are excluded for $m_\PSA=600\GeV$ and $m_\Pa=200\GeV$, assuming $\tan\beta=1$.
Also excluded are $\tan\beta$ values between 0.5 and 2.0 (1.6) for $m_\Pa=100$ (150)\GeV and $m_\PSA=600\GeV$, given $\sin\theta=0.35$.
These are the first experimental limits on the 2HDM+\Pa model.

\begin{figure*}[htbp]
  \centering
  \includegraphics[width=0.61\textwidth]{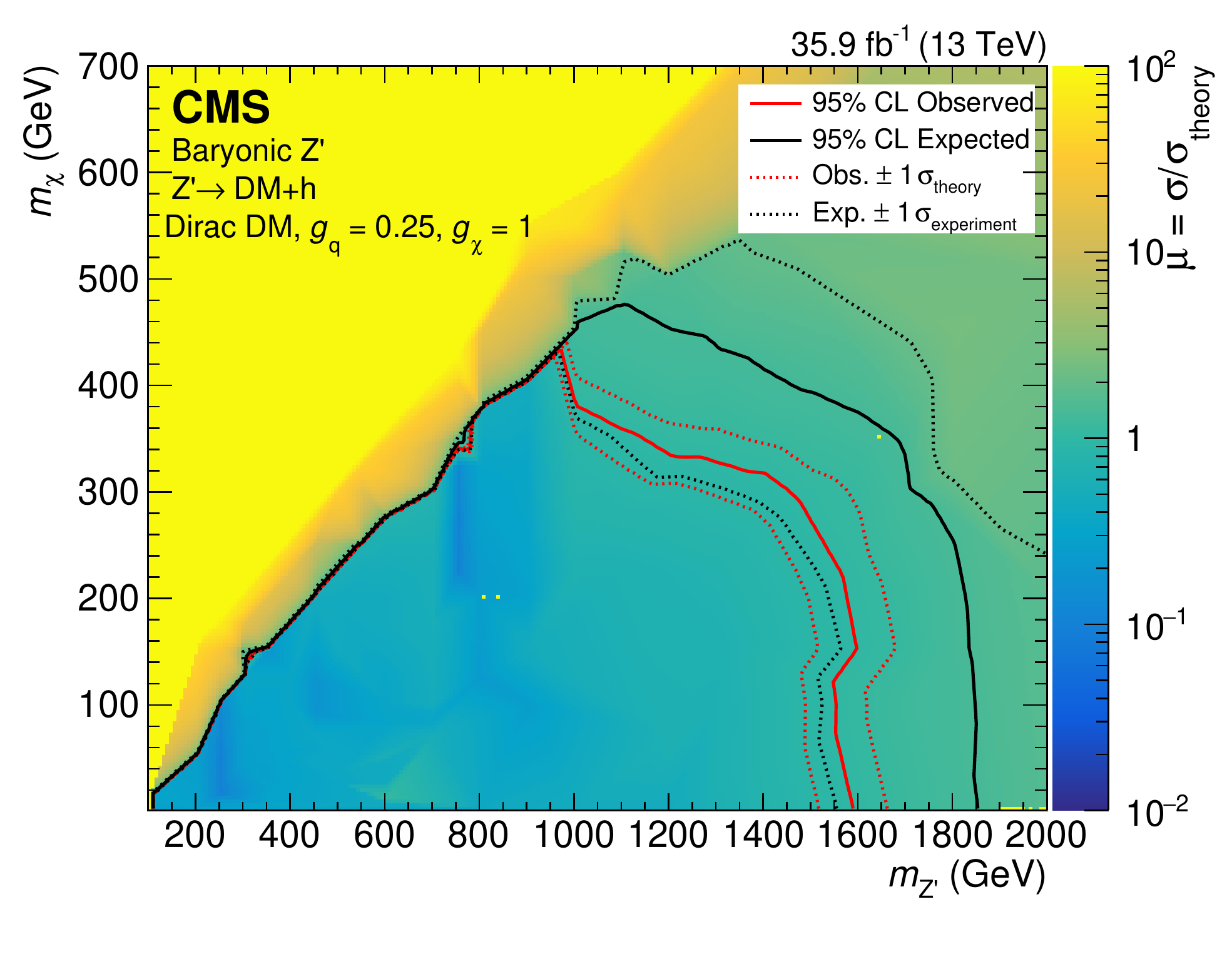}
  \caption{Upper limits at 95\% \CL on the signal strength modifier, defined as $\mu=\sigma/\sigma_\text{theory}$, where $\sigma_\text{theory}$ is the predicted production cross section of DM candidates in association with a Higgs boson and $\sigma$ is the upper limit on the observed cross section. Limits are shown for the baryonic \cPZpr\ model  as a function of $m_{\cPZpr}$ and $m_\chi$. Mediators of up to 1.6\TeV are excluded for a DM mass of 1\GeV. Masses of the DM particle itself are excluded up to 430\GeV for a \cPZpr\ mass of 960\GeV.}
  \label{fig:limits}
\end{figure*}

Figure~\ref{fig:limits} shows the expected and observed exclusion range as a function of $m_\cPZpr$ and $m_{\chi}$ for the baryonic \cPZpr\ model.
For a DM mass of 1\GeV, masses $m_{\cPZpr}<1.6\TeV$ are excluded.
The expected exclusion boundary is 1.85\TeV.
Masses for the DM particles of up to 430\GeV are excluded for a 960\GeV \cPZpr\ mass.
These are the most stringent limits on this model so far.

\begin{figure*}
  \centering
  \includegraphics[width=0.61\textwidth]{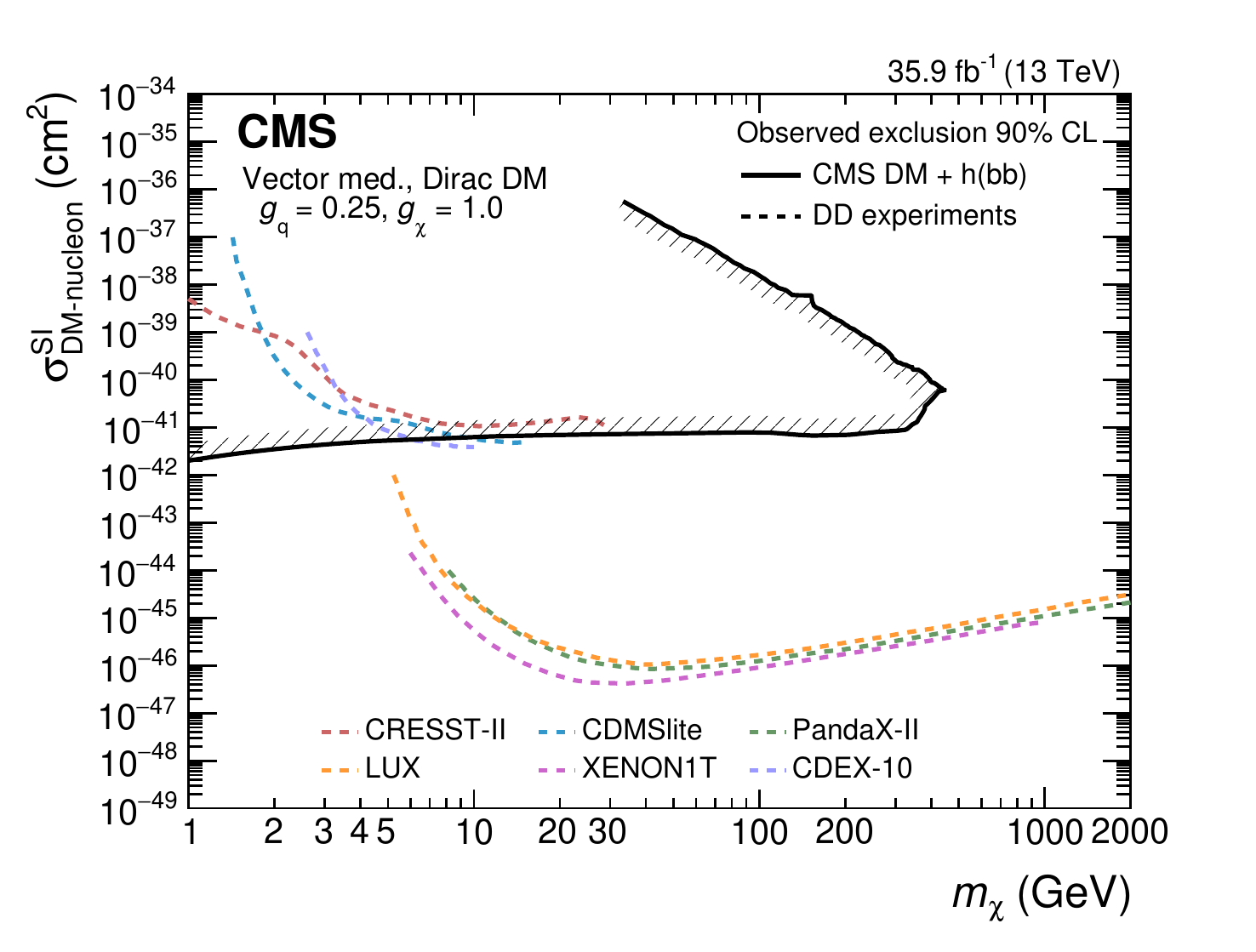}
  \caption{The 90\% \CL exclusion limits on the DM-nucleon SI scattering cross section as a function of $m_{\chi}$.
Results for the baryonic \cPZpr\ model obtained in this analysis are compared with those from a selection of direct detection (DD) experiments.
The latter exclude the regions above the curves.
Limits from CRESST-II~\cite{CresstII}, CDMSlite~\cite{CDMSLite}, LUX~\cite{LUX}, XENON-1T~\cite{XENON1T}, PandaX-II~\cite{PandaxII}, and CDEX-10~\cite{Jiang:2018pic} are shown.}
  \label{fig:limitsdd}
\end{figure*}

To compare results with DM direct detection experiments, limits from the baryonic \cPZpr\ model are presented in terms of a spin-independent (SI) cross section \SigSI for DM scattering off a nucleus.
Following the recommendation of Ref.~\cite{presentDM}, the value of $\sigma_\mathrm{SI}$ is determined by the equation:
\begin{equation}
\sigma_\text{SI} = \frac{f^2(g_{\cPq})g^2_{\chi}\mu^2_{\mathrm{n}\chi}}{\pi m^4_{\mathrm{med}}},
\end{equation}
where $\mu_{\mathrm{n}\chi}$ is the reduced mass of the DM-nucleon system, $f(g_{\cPq})$ is the mediator-nucleon coupling, which depends on the mediator coupling to SM quarks $g_{\cPq}$, $g_{\chi}$ is the mediator coupling to SM particles, and $m_{\text{med}}$ is the mass of the mediator.
The resulting \SigSI limits as a function of the DM mass are shown in Fig.~\ref{fig:limitsdd}.
Under the assumptions made for the baryonic \cPZpr\ model, these limits on the DM-nucleon SI cross section are the most stringent to date for $m_\chi < 5\GeV$.

\section{Summary}

A search for dark matter (DM) produced in association with a Higgs boson decaying to a pair of bottom quarks in a sample of proton-proton collision data corresponding to 35.9\fbinv is presented.
No significant deviation from the predictions of the standard model is observed, and 95\% \CL upper limits on the production cross sections predicted by a type-2 two-Higgs doublet model extended by an additional light pseudoscalar boson \Pa (2HDM+\Pa) and by the baryonic \cPZpr\ model are established.
These limits constitute the most stringent exclusions from collider experiments placed on the parameters of these models to date.
The 2HDM+\Pa model is probed experimentally for the first time.
For the nominal choice of the mixing angles $\sin\theta$ and $\tan\beta$ in the 2HDM+\Pa model, the search excludes masses $500<m_\PSA<900\GeV$ (where \PSA is the heavy pseudoscalar boson) assuming $m_\Pa=150\GeV$.
Scanning over $\sin\theta$ with $\tan\beta$ = 1, we exclude $0.35<\sin\theta<0.75$ for $m_\PSA=600\GeV$ and $m_\Pa=200\GeV$.
Finally, $\tan\beta$ values between 0.5 and 2.0 (1.6) are excluded for $m_\PSA=600\GeV$ and $m_\Pa=100$ (150)\GeV and $\sin\theta = 0.35$.
In all 2HDM+\Pa interpretations, a DM mass of $m_\chi=10\GeV$ is assumed.
For the baryonic \cPZpr\ model, we exclude \cPZpr\ boson masses up to 1.6\TeV for a DM mass of 1\GeV, and DM masses up to 430\GeV for a \cPZpr\ boson mass of 960\GeV.
The reinterpretation of the results for the baryonic \cPZpr\ model in terms of an SI nucleon scattering cross section yields a higher sensitivity for $m_\chi<5\GeV$ than existing results from direct detection experiments, under the assumptions imposed by the model.

\begin{acknowledgments}
We congratulate our colleagues in the CERN accelerator departments for the excellent performance of the LHC and thank the technical and administrative staffs at CERN and at other CMS institutes for their contributions to the success of the CMS effort. In addition, we gratefully acknowledge the computing centers and personnel of the Worldwide LHC Computing Grid for delivering so effectively the computing infrastructure essential to our analyses. Finally, we acknowledge the enduring support for the construction and operation of the LHC and the CMS detector provided by the following funding agencies: BMBWF and FWF (Austria); FNRS and FWO (Belgium); CNPq, CAPES, FAPERJ, FAPERGS, and FAPESP (Brazil); MES (Bulgaria); CERN; CAS, MoST, and NSFC (China); COLCIENCIAS (Colombia); MSES and CSF (Croatia); RPF (Cyprus); SENESCYT (Ecuador); MoER, ERC IUT, and ERDF (Estonia); Academy of Finland, MEC, and HIP (Finland); CEA and CNRS/IN2P3 (France); BMBF, DFG, and HGF (Germany); GSRT (Greece); NKFIA (Hungary); DAE and DST (India); IPM (Iran); SFI (Ireland); INFN (Italy); MSIP and NRF (Republic of Korea); MES (Latvia); LAS (Lithuania); MOE and UM (Malaysia); BUAP, CINVESTAV, CONACYT, LNS, SEP, and UASLP-FAI (Mexico); MOS (Montenegro); MBIE (New Zealand); PAEC (Pakistan); MSHE and NSC (Poland); FCT (Portugal); JINR (Dubna); MON, RosAtom, RAS, RFBR, and NRC KI (Russia); MESTD (Serbia); SEIDI, CPAN, PCTI, and FEDER (Spain); MOSTR (Sri Lanka); Swiss Funding Agencies (Switzerland); MST (Taipei); ThEPCenter, IPST, STAR, and NSTDA (Thailand); TUBITAK and TAEK (Turkey); NASU and SFFR (Ukraine); STFC (United Kingdom); DOE and NSF (USA).

\hyphenation{Rachada-pisek} Individuals have received support from the Marie-Curie program and the European Research Council and Horizon 2020 Grant, contract No. 675440 (European Union); the Leventis Foundation; the A. P. Sloan Foundation; the Alexander von Humboldt Foundation; the Belgian Federal Science Policy Office; the Fonds pour la Formation \`a la Recherche dans l'Industrie et dans l'Agriculture (FRIA-Belgium); the Agentschap voor Innovatie door Wetenschap en Technologie (IWT-Belgium); the F.R.S.-FNRS and FWO (Belgium) under the ``Excellence of Science - EOS" - be.h project n. 30820817; the Ministry of Education, Youth and Sports (MEYS) of the Czech Republic; the Lend\"ulet (``Momentum") Program and the J\'anos Bolyai Research Scholarship of the Hungarian Academy of Sciences, the New National Excellence Program \'UNKP, the NKFIA research grants 123842, 123959, 124845, 124850 and 125105 (Hungary); the Council of Science and Industrial Research, India; the HOMING PLUS program of the Foundation for Polish Science, cofinanced from European Union, Regional Development Fund, the Mobility Plus program of the Ministry of Science and Higher Education, the National Science Center (Poland), contracts Harmonia 2014/14/M/ST2/00428, Opus 2014/13/B/ST2/02543, 2014/15/B/ST2/03998, and 2015/19/B/ST2/02861, Sonata-bis 2012/07/E/ST2/01406; the National Priorities Research Program by Qatar National Research Fund; the Programa Estatal de Fomento de la Investigaci{\'o}n Cient{\'i}fica y T{\'e}cnica de Excelencia Mar\'{\i}a de Maeztu, grant MDM-2015-0509 and the Programa Severo Ochoa del Principado de Asturias; the Thalis and Aristeia programs cofinanced by EU-ESF and the Greek NSRF; the Rachadapisek Sompot Fund for Postdoctoral Fellowship, Chulalongkorn University and the Chulalongkorn Academic into Its 2nd Century Project Advancement Project (Thailand); the Welch Foundation, contract C-1845; and the Weston Havens Foundation (USA). \end{acknowledgments}

\clearpage
\bibliography{auto_generated}

\cleardoublepage \appendix\section{The CMS Collaboration \label{app:collab}}\begin{sloppypar}\hyphenpenalty=5000\widowpenalty=500\clubpenalty=5000\input{EXO-16-050-authorlist.tex}\end{sloppypar}
\end{document}

%% file: EXO-16-050-authorlist.tex
\vskip\cmsinstskip
\textbf{Yerevan Physics Institute, Yerevan, Armenia}\\*[0pt]
A.M.~Sirunyan, A.~Tumasyan
\vskip\cmsinstskip
\textbf{Institut f\"{u}r Hochenergiephysik, Wien, Austria}\\*[0pt]
W.~Adam, F.~Ambrogi, E.~Asilar, T.~Bergauer, J.~Brandstetter, M.~Dragicevic, J.~Er\"{o}, A.~Escalante~Del~Valle, M.~Flechl, R.~Fr\"{u}hwirth\cmsAuthorMark{1}, V.M.~Ghete, J.~Hrubec, M.~Jeitler\cmsAuthorMark{1}, N.~Krammer, I.~Kr\"{a}tschmer, D.~Liko, T.~Madlener, I.~Mikulec, N.~Rad, H.~Rohringer, J.~Schieck\cmsAuthorMark{1}, R.~Sch\"{o}fbeck, M.~Spanring, D.~Spitzbart, A.~Taurok, W.~Waltenberger, J.~Wittmann, C.-E.~Wulz\cmsAuthorMark{1}, M.~Zarucki
% ...
\vskip\cmsinstskip
% ...
1:  Also at Vienna University of Technology, Vienna, Austria\\
% ...


%% file: EXO-16-050_temp.bbl
\providecommand{\href}[2]{#2}\begingroup\raggedright\begin{thebibliography}{10}%
\makeatletter
\providecommand{\hrefCMSnoop }[0]{\@secondoftwo}%
\makeatother
\providecommand{\doi}{\texttt{doi:}\begingroup \urlstyle{tt}\Url}

\bibitem{dm1}
\hrefCMSnoop {}{G.~Bertone, D.~Hooper, and J.~Silk, ``Particle dark matter:
  Evidence, candidates and constraints'',} \textit{ Phys. Rept.} \textbf{ 405}
  (2005) 279,
  \href{http://dx.doi.org/10.1016/j.physrep.2004.08.031}{\doi{10.1016/j.physrep.2004.08.031}},
  \href{http://www.arXiv.org/abs/hep-ph/0404175}{\texttt{arXiv:hep-ph/0404175}}.

\bibitem{dm2}
\hrefCMSnoop {}{J.~L. Feng, ``Dark matter candidates from particle physics and
  methods of detection'',} \textit{ Ann. Rev. Astron. Astrophys.} \textbf{ 48}
  (2010) 495,
  \href{http://dx.doi.org/10.1146/annurev-astro-082708-101659}{\doi{10.1146/annurev-astro-082708-101659}},
\href{http://www.arXiv.org/abs/1003.0904}{\texttt{arXiv:1003.0904}}.

\bibitem{dm3}
\hrefCMSnoop {}{T.~A. Porter, R.~P. Johnson, and P.~W. Graham, ``Dark matter
  searches with astroparticle data'',} \textit{ Ann. Rev. Astron. Astrophys.}
  \textbf{ 49} (2011) 155,
  \href{http://dx.doi.org/10.1146/annurev-astro-081710-102528}{\doi{10.1146/annurev-astro-081710-102528}},
\href{http://www.arXiv.org/abs/1104.2836}{\texttt{arXiv:1104.2836}}.

\bibitem{planck}
\hrefCMSnoop {}{{Planck} Collaboration, ``{Planck 2015 results. XIII.
  Cosmological parameters}'',} \textit{ Astron. Astrophys.} \textbf{ 594}
  (2016) A13,
  \href{http://dx.doi.org/10.1051/0004-6361/201525830}{\doi{10.1051/0004-6361/201525830}},
\href{http://www.arXiv.org/abs/1502.01589}{\texttt{arXiv:1502.01589}}.

\bibitem{HiggsObs_ATLAS}
\hrefCMSnoop {}{{ATLAS Collaboration}, ``Observation of a new particle in the
  search for the standard model {Higgs} boson with the {ATLAS} detector at the
  {LHC}'',} \textit{ Phys. Lett. B} \textbf{ 716} (2012) 1,
  \href{http://dx.doi.org/10.1016/j.physletb.2012.08.020}{\doi{10.1016/j.physletb.2012.08.020}},
  \href{http://www.arXiv.org/abs/1207.7214}{\texttt{arXiv:1207.7214}}.

\bibitem{HiggsObs_CMS}
\hrefCMSnoop {}{{CMS Collaboration}, ``Observation of a new boson at a mass of
  125 {GeV} with the {CMS} experiment at the {LHC}'',} \textit{ Phys. Lett. B}
  \textbf{ 716} (2012) 30,
  \href{http://dx.doi.org/10.1016/j.physletb.2012.08.021}{\doi{10.1016/j.physletb.2012.08.021}},
  \href{http://www.arXiv.org/abs/1207.7235}{\texttt{arXiv:1207.7235}}.

\bibitem{HiggsObs_CMS_Long}
\hrefCMSnoop {}{{CMS Collaboration}, ``{Observation of a new boson with mass
  near 125\,GeV in pp collisions at $\sqrt{s} = 7$ and $8\TeV$}'',} \textit{
  JHEP} \textbf{ 06} (2013) 81,
  \href{http://dx.doi.org/10.1007/JHEP06(2013)081}{\doi{10.1007/JHEP06(2013)081}},
  \href{http://www.arXiv.org/abs/1303.4571}{\texttt{arXiv:1303.4571}}.

\bibitem{monoHiggs3}
\hrefCMSnoop {}{A.~A. Petrov and W.~Shepherd, ``{Searching for dark matter at
  LHC with mono-Higgs production}'',} \textit{ Phys. Lett. B} \textbf{ 730}
  (2014) 178,
  \href{http://dx.doi.org/10.1016/j.physletb.2014.01.051}{\doi{10.1016/j.physletb.2014.01.051}},
\href{http://www.arXiv.org/abs/1311.1511}{\texttt{arXiv:1311.1511}}.

\bibitem{2HDM}
\hrefCMSnoop {}{A.~Berlin, T.~Lin, and L.-T. Wang, ``{Mono-Higgs detection of
  dark matter at the LHC}'',} \textit{ JHEP} \textbf{ 06} (2014) 078,
  \href{http://dx.doi.org/10.1007/JHEP06(2014)078}{\doi{10.1007/JHEP06(2014)078}},
\href{http://www.arXiv.org/abs/1402.7074}{\texttt{arXiv:1402.7074}}.

\bibitem{PhysRevD.89.075017}
L.~Carpenter\hrefCMSnoop {}{ {et~al.}, ``Mono-{H}iggs-boson: A new collider
  probe of dark matter'',} \textit{ Phys. Rev. D} \textbf{ 89} (2014) 075017,
  \href{http://dx.doi.org/10.1103/PhysRevD.89.075017}{\doi{10.1103/PhysRevD.89.075017}},
  \href{http://www.arXiv.org/abs/1312.2592}{\texttt{arXiv:1312.2592}}.

\bibitem{CMSdetector}
\hrefCMSnoop {}{{CMS Collaboration}, ``The {CMS} experiment at the {CERN
  LHC}'',} \textit{ JINST} \textbf{ 3} (2008) 08004,
  \href{http://dx.doi.org/10.1088/1748-0221/3/08/S08004}{\doi{10.1088/1748-0221/3/08/S08004}}.

\bibitem{PhysRevLett.115.131801}
\hrefCMSnoop {}{{ATLAS Collaboration}, ``Search for dark matter in events with
  missing transverse momentum and a {Higgs} boson decaying to two photons in pp
  collisions at $\sqrt{s}=8\,\mathrm{TeV}$ with the {ATLAS} detector'',}
  \textit{ Phys. Rev. Lett.} \textbf{ 115} (2015) 131801,
  \href{http://dx.doi.org/10.1103/PhysRevLett.115.131801}{\doi{10.1103/PhysRevLett.115.131801}}.

\bibitem{PhysRevLett.119.181804}
\hrefCMSnoop {}{{ATLAS Collaboration}, ``Search for dark matter produced in
  association with a {Higgs} boson decaying to $\text{b}\overline{\text{b}}$
  using $36\,{\mathrm{fb}}^{\ensuremath{-}1}$ of pp collisions at
  $\sqrt{s}=13\,\mathrm{TeV}$ with the {ATLAS} detector'',} \textit{ Phys. Rev.
  Lett.} \textbf{ 119} (2017) 181804,
  \href{http://dx.doi.org/10.1103/PhysRevLett.119.181804}{\doi{10.1103/PhysRevLett.119.181804}}.

\bibitem{1807.02826}
\hrefCMSnoop {}{{CMS Collaboration}, ``{Search for heavy resonances decaying
  into a vector boson and a Higgs boson in final states with charged leptons,
  neutrinos and b quarks at $\sqrt{s} = 13\TeV$}'',} (2018).
  \href{http://www.arXiv.org/abs/1807.02826}{\texttt{arXiv:1807.02826}}.
Submitted to \textit{JHEP}.

\bibitem{Bauer2017}
\hrefCMSnoop {}{M.~Bauer, U.~Haisch, and F.~Kahlhoefer, ``{Simplified dark
  matter models with two Higgs doublets: I. Pseudoscalar mediators}'',}
  \textit{ JHEP} \textbf{ 05} (2017) 138,
  \href{http://dx.doi.org/10.1007/JHEP05(2017)138}{\doi{10.1007/JHEP05(2017)138}},
  \href{http://www.arXiv.org/abs/1701.07427}{\texttt{arXiv:1701.07427}}.

\bibitem{Khachatryan:2016vau}
\hrefCMSnoop {}{{ATLAS and CMS Collaborations}, ``{Measurements of the Higgs
  boson production and decay rates and constraints on its couplings from a
  combined ATLAS and CMS analysis of the LHC pp collision data at $ \sqrt{s}=7
  $ and $8\TeV$}'',} \textit{ JHEP} \textbf{ 08} (2016) 045,
  \href{http://dx.doi.org/10.1007/JHEP08(2016)045}{\doi{10.1007/JHEP08(2016)045}},
\href{http://www.arXiv.org/abs/1606.02266}{\texttt{arXiv:1606.02266}}.

\bibitem{Sirunyan:2017hnk}
\hrefCMSnoop {}{{CMS Collaboration}, ``{Search for associated production of
  dark matter with a Higgs boson decaying to $ \mathrm{b}\overline{\mathrm{b}}
  $ or $\gamma \gamma$ at $ \sqrt{s}=13\TeV$}'',} \textit{ JHEP} \textbf{ 10}
  (2017) 180,
  \href{http://dx.doi.org/10.1007/JHEP10(2017)180}{\doi{10.1007/JHEP10(2017)180}},
\href{http://www.arXiv.org/abs/1703.05236}{\texttt{arXiv:1703.05236}}.

\bibitem{Abe:2018bpo}
\hrefCMSnoop {}{{LHC Dark Matter Working Group} Collaboration, ``{LHC Dark
  Matter Working Group: Next-generation spin-0 dark matter models}'',} (2018).
\href{http://www.arXiv.org/abs/1810.09420}{\texttt{arXiv:1810.09420}}.

\bibitem{Abercrombie:2015wmb}
\hrefCMSnoop {}{D.~Abercrombie {et~al.}, ``Dark matter benchmark models for
  early {LHC} run-2 searches: Report of the {ATLAS}/{CMS} dark matter forum'',}
  (2015).
\href{http://www.arXiv.org/abs/1507.00966}{\texttt{arXiv:1507.00966}}.

\bibitem{1748-0221-12-01-P01020}
\hrefCMSnoop {}{{CMS Collaboration}, ``The {CMS} trigger system'',} \textit{
  JINST} \textbf{ 12} (2017) P01020,
  \href{http://dx.doi.org/10.1088/1748-0221/12/01/P01020}{\doi{10.1088/1748-0221/12/01/P01020}},
  \href{http://www.arXiv.org/abs/1609.02366}{\texttt{arXiv:1609.02366}}.

\bibitem{amcatnlo}
J.~Alwall\hrefCMSnoop {}{ {et~al.}, ``{The automated computation of tree-level
  and next-to-leading order differential cross sections, and their matching to
  parton shower simulations}'',} \textit{ JHEP} \textbf{ 07} (2014) 079,
  \href{http://dx.doi.org/10.1007/JHEP07(2014)079}{\doi{10.1007/JHEP07(2014)079}},
\href{http://www.arXiv.org/abs/1405.0301}{\texttt{arXiv:1405.0301}}.

\bibitem{Nason:2004rx}
\hrefCMSnoop {}{P.~Nason, ``{A new method for combining NLO QCD with shower
  Monte Carlo algorithms}'',} \textit{ JHEP} \textbf{ 11} (2004) 040,
  \href{http://dx.doi.org/10.1088/1126-6708/2004/11/040}{\doi{10.1088/1126-6708/2004/11/040}},
\href{http://www.arXiv.org/abs/hep-ph/0409146}{\texttt{arXiv:hep-ph/0409146}}.

\bibitem{Frixione:2007vw}
\hrefCMSnoop {}{S.~Frixione, P.~Nason, and C.~Oleari, ``{Matching NLO QCD
  computations with parton shower simulations: the POWHEG method}'',} \textit{
  JHEP} \textbf{ 11} (2007) 070,
  \href{http://dx.doi.org/10.1088/1126-6708/2007/11/070}{\doi{10.1088/1126-6708/2007/11/070}},
\href{http://www.arXiv.org/abs/0709.2092}{\texttt{arXiv:0709.2092}}.

\bibitem{Alioli:2010xd}
\hrefCMSnoop {}{S.~Alioli, P.~Nason, C.~Oleari, and E.~Re, ``{A general
  framework for implementing NLO calculations in shower Monte Carlo programs:
  the POWHEG BOX}'',} \textit{ JHEP} \textbf{ 06} (2010) 043,
  \href{http://dx.doi.org/10.1007/JHEP06(2010)043}{\doi{10.1007/JHEP06(2010)043}},
  \href{http://www.arXiv.org/abs/1002.2581}{\texttt{arXiv:1002.2581}}.

\bibitem{ttbarNNLO}
\hrefCMSnoop {}{M.~Czakon, P.~Fiedler, and A.~Mitov, ``Total top-quark
  pair-production cross section at hadron colliders through $o(\alpha^4_s)$'',}
  \textit{ Phys. Rev. Lett.} \textbf{ 110} (2013) 252004,
  \href{http://dx.doi.org/10.1103/PhysRevLett.110.252004}{\doi{10.1103/PhysRevLett.110.252004}},
\href{http://www.arXiv.org/abs/1303.6254}{\texttt{arXiv:1303.6254}}.

\bibitem{mlm}
\hrefCMSnoop {}{M.~L. Mangano, M.~Moretti, F.~Piccinini, and M.~Treccani,
  ``Matching matrix elements and shower evolution for top-quark production in
  hadronic collisions'',} \textit{ JHEP} \textbf{ 01} (2007) 013,
  \href{http://dx.doi.org/10.1088/1126-6708/2007/01/013}{\doi{10.1088/1126-6708/2007/01/013}},
  \href{http://www.arXiv.org/abs/hep-ph/0611129}{\texttt{arXiv:hep-ph/0611129}}.

\bibitem{fxfx}
\hrefCMSnoop {}{R.~Frederix and S.~Frixione, ``{Merging meets matching in
  MC@NLO}'',} \textit{ JHEP} \textbf{ 12} (2012) 061,
  \href{http://dx.doi.org/10.1007/JHEP12(2012)061}{\doi{10.1007/JHEP12(2012)061}},
\href{http://www.arXiv.org/abs/1209.6215}{\texttt{arXiv:1209.6215}}.

\bibitem{Kuhn:2005gv}
\hrefCMSnoop {}{J.~H. Kuhn, A.~Kulesza, S.~Pozzorini, and M.~Schulze,
  ``{Electroweak corrections to hadronic photon production at large transverse
  momenta}'',} \textit{ JHEP} \textbf{ 03} (2006) 059,
  \href{http://dx.doi.org/10.1088/1126-6708/2006/03/059}{\doi{10.1088/1126-6708/2006/03/059}},
\href{http://www.arXiv.org/abs/hep-ph/0508253}{\texttt{arXiv:hep-ph/0508253}}.

\bibitem{Kallweit:2015fta}
S.~Kallweit\href
  {http://inspirehep.net/record/1372103/files/arXiv:1505.05704.pdf}{ {et~al.},
  ``{NLO QCD+EW automation and precise predictions for V+multijet
  production}'',} in \textit{ {50th Rencontres de Moriond on QCD and High
  Energy Interactions La Thuile, Italy, March 21-28, 2015}}.
\newblock 2015.
\newblock
\href{http://www.arXiv.org/abs/1505.05704}{\texttt{arXiv:1505.05704}}.
\newblock

\bibitem{Kallweit:2015dum}
S.~Kallweit\hrefCMSnoop {}{ {et~al.}, ``{NLO QCD+EW predictions for V+jets
  including off-shell vector-boson decays and multijet merging}'',} \textit{
  JHEP} \textbf{ 04} (2016) 021,
  \href{http://dx.doi.org/10.1007/JHEP04(2016)021}{\doi{10.1007/JHEP04(2016)021}},
\href{http://www.arXiv.org/abs/1511.08692}{\texttt{arXiv:1511.08692}}.

\bibitem{Sjostrand:2014zea}
T.~Sj{\"o}strand\hrefCMSnoop {}{ {et~al.}, ``{An Introduction to PYTHIA
  8.2}'',} \textit{ Comput. Phys. Commun.} \textbf{ 191} (2015) 159,
  \href{http://dx.doi.org/10.1016/j.cpc.2015.01.024}{\doi{10.1016/j.cpc.2015.01.024}},
\href{http://www.arXiv.org/abs/1410.3012}{\texttt{arXiv:1410.3012}}.

\bibitem{MCFM}
\hrefCMSnoop {}{J.~M. Campbell, R.~K. Ellis, and C.~Williams, ``{Vector boson
  pair production at the LHC}'',} \textit{ JHEP} \textbf{ 07} (2011) 018,
  \href{http://dx.doi.org/10.1007/JHEP07(2011)018}{\doi{10.1007/JHEP07(2011)018}},
\href{http://www.arXiv.org/abs/1105.0020}{\texttt{arXiv:1105.0020}}.

\bibitem{Ball:2014uwa}
\hrefCMSnoop {}{{NNPDF} Collaboration, ``{Parton distributions for the LHC Run
  II}'',} \textit{ JHEP} \textbf{ 04} (2015) 040,
  \href{http://dx.doi.org/10.1007/JHEP04(2015)040}{\doi{10.1007/JHEP04(2015)040}},
\href{http://www.arXiv.org/abs/1410.8849}{\texttt{arXiv:1410.8849}}.

\bibitem{ue1}
\hrefCMSnoop {}{{CMS Collaboration}, ``Event generator tunes obtained from
  underlying event and multiparton scattering measurements'',} \textit{ Eur.
  Phys. J. C} \textbf{ 76} (2016) 155,
  \href{http://dx.doi.org/10.1140/epjc/s10052-016-3988-x}{\doi{10.1140/epjc/s10052-016-3988-x}},
  \href{http://www.arXiv.org/abs/1512.00815}{\texttt{arXiv:1512.00815}}.

\bibitem{ue2}
\hrefCMSnoop {}{P.~Skands, S.~Carrazza, and J.~Rojo, ``{Tuning PYTHIA 8.1: the
  Monash 2013 tune}'',} \textit{ Eur. Phys. J. C} \textbf{ 74} (2014) 3024,
  \href{http://dx.doi.org/10.1140/epjc/s10052-014-3024-y}{\doi{10.1140/epjc/s10052-014-3024-y}},
  \href{http://www.arXiv.org/abs/1404.5630}{\texttt{arXiv:1404.5630}}.

\bibitem{geant4}
\hrefCMSnoop {}{{GEANT4} Collaboration, ``{\GEANTfour}---a simulation
  toolkit'',} \textit{ Nucl. Instrum. Meth. A} \textbf{ 506} (2003) 250,
\href{http://dx.doi.org/10.1016/S0168-9002(03)01368-8}{\doi{10.1016/S0168-9002(03)01368-8}}.

\bibitem{Cacciari:2008gp}
\hrefCMSnoop {}{M.~Cacciari, G.~P. Salam, and G.~Soyez, ``The anti-$\kt$ jet
  clustering algorithm'',} \textit{ JHEP} \textbf{ 04} (2008) 063,
  \href{http://dx.doi.org/10.1088/1126-6708/2008/04/063}{\doi{10.1088/1126-6708/2008/04/063}},
  \href{http://www.arXiv.org/abs/0802.1189}{\texttt{arXiv:0802.1189}}.

\bibitem{Cacciari:2011ma}
\hrefCMSnoop {}{M.~Cacciari, G.~P. Salam, and G.~Soyez, ``{FastJet user
  manual}'',} \textit{ Eur. Phys. J. C} \textbf{ 72} (2012) 1896,
  \href{http://dx.doi.org/10.1140/epjc/s10052-012-1896-2}{\doi{10.1140/epjc/s10052-012-1896-2}},
\href{http://www.arXiv.org/abs/1111.6097}{\texttt{arXiv:1111.6097}}.

\bibitem{Sirunyan:2017ulk}
\hrefCMSnoop {}{{CMS Collaboration}, ``{Particle-flow reconstruction and global
  event description with the CMS detector}'',} \textit{ JINST} \textbf{ 12}
  (2017) 10003,
  \href{http://dx.doi.org/10.1088/1748-0221/12/10/P10003}{\doi{10.1088/1748-0221/12/10/P10003}},
\href{http://www.arXiv.org/abs/1706.04965}{\texttt{arXiv:1706.04965}}.

\bibitem{cajets}
\href {http://cds.cern.ch/record/1194489}{{CMS Collaboration}, ``{A
  Cambridge-Aachen (C-A) based jet algorithm for boosted top-jet tagging}'',}
  CMS Physics Analysis Summary CMS-PAS-JME-09-001, 2009.

\bibitem{puppi}
\hrefCMSnoop {}{D.~Berteloni, P.~Harris, M.~Low, and N.~Tran, ``Pileup per
  particle identification'',} \textit{ JHEP} \textbf{ 59} (2014) 059,
  \href{http://dx.doi.org/10.1007/JHEP10(2014)059}{\doi{10.1007/JHEP10(2014)059}},
  \href{http://www.arXiv.org/abs/1407.6013}{\texttt{arXiv:1407.6013}}.

\bibitem{jec}
\hrefCMSnoop {}{{CMS Collaboration}, ``{Jet energy scale and resolution in the
  CMS experiment in pp collisions at $8\TeV$}'',} \textit{ JINST} \textbf{ 12}
  (2017) 02014,
  \href{http://dx.doi.org/10.1088/1748-0221/12/02/P02014}{\doi{10.1088/1748-0221/12/02/P02014}},
  \href{http://www.arXiv.org/abs/1607.03663}{\texttt{arXiv:1607.03663}}.

\bibitem{msd}
\hrefCMSnoop {}{A.~J. Larkoski, S.~Marzani, G.~Soyez, and J.~Thaler, ``Soft
  drop'',} \textit{ JHEP} \textbf{ 05} (2014) 146,
  \href{http://dx.doi.org/10.1007/JHEP05(2014)146}{\doi{10.1007/JHEP05(2014)146}},
  \href{http://www.arXiv.org/abs/1402.2657}{\texttt{arXiv:1402.2657}}.

\bibitem{Sirunyan:2017ezt}
\hrefCMSnoop {}{{CMS Collaboration}, ``{Identification of heavy-flavour jets
  with the CMS detector in pp collisions at $13\TeV$}'',} \textit{ JINST}
  \textbf{ 13} (2018) 05011,
  \href{http://dx.doi.org/10.1088/1748-0221/13/05/P05011}{\doi{10.1088/1748-0221/13/05/P05011}},
\href{http://www.arXiv.org/abs/1712.07158}{\texttt{arXiv:1712.07158}}.

\bibitem{ecf}
\hrefCMSnoop {}{I.~Moult, L.~Necib, and J.~Thaler, ``New angles on energy
  correlation functions'',} \textit{ JHEP} \textbf{ 12} (2016) 153,
  \href{http://dx.doi.org/10.1007/JHEP12(2016)153}{\doi{10.1007/JHEP12(2016)153}},
\href{http://www.arXiv.org/abs/1609.07483}{\texttt{arXiv:1609.07483}}.

\bibitem{ddt}
J.~Dolen\hrefCMSnoop {}{ {et~al.}, ``{Thinking outside the ROCs: Designing
  decorrelated taggers (DDT) for jet substructure}'',} \textit{ JHEP} \textbf{
  05} (2016) 156,
  \href{http://dx.doi.org/10.1007/JHEP05(2016)156}{\doi{10.1007/JHEP05(2016)156}},
\href{http://www.arXiv.org/abs/1603.00027}{\texttt{arXiv:1603.00027}}.

\bibitem{Khachatryan:2015hwa}
\hrefCMSnoop {}{{CMS Collaboration}, ``Performance of electron reconstruction
  and selection with the {CMS} detector in proton-proton collisions at
  $\sqrt{s} = 8$\,tev'',} \textit{ JINST} \textbf{ 10} (2015) 06005,
  \href{http://dx.doi.org/10.1088/1748-0221/10/06/P06005}{\doi{10.1088/1748-0221/10/06/P06005}},
\href{http://www.arXiv.org/abs/1502.02701}{\texttt{arXiv:1502.02701}}.

\bibitem{CMSMuonJINST}
\hrefCMSnoop {}{{CMS Collaboration}, ``{Performance of CMS muon reconstruction
  in pp collision events at $\sqrt{s} = 7\TeV$}'',} \textit{ JINST} \textbf{ 7}
  (2012) 10002,
  \href{http://dx.doi.org/10.1088/1748-0221/7/10/P10002}{\doi{10.1088/1748-0221/7/10/P10002}},
\href{http://www.arXiv.org/abs/1206.4071}{\texttt{arXiv:1206.4071}}.

\bibitem{CMSTauJINST}
\hrefCMSnoop {}{{CMS Collaboration}, ``{Reconstruction and identification of
  $\tau$ lepton decays to hadrons and $\nu_{\tau}$ at CMS}'',} \textit{ JINST}
  \textbf{ 11} (2016) 01019,
  \href{http://dx.doi.org/10.1088/1748-0221/11/01/P01019}{\doi{10.1088/1748-0221/11/01/P01019}},
\href{http://www.arXiv.org/abs/1510.07488}{\texttt{arXiv:1510.07488}}.

\bibitem{Chatrchyan:2013sba}
\hrefCMSnoop {}{{CMS Collaboration}, ``{Performance of the CMS muon detector
  and muon reconstruction with proton-proton collisions $\sqrt{s} =
  13\TeV$}'',} \textit{ JINST} \textbf{ 13} (2018) 06015,
  \href{http://dx.doi.org/10.1088/1748-0221/13/06/P06015}{\doi{10.1088/1748-0221/13/06/P06015}},
\href{http://www.arXiv.org/abs/1804.04528}{\texttt{arXiv:1804.04528}}.

\bibitem{CMS-PAS-JME-16-004}
\href {http://cds.cern.ch/record/2205284?ln=en}{{CMS Collaboration},
  ``{Performance of missing energy reconstruction in $13\TeV$ pp collision data
  using the CMS detector}'',} CMS Physics Analysis Summary CMS-PAS-JME-16-004,
  2016.

\bibitem{RooStats}
L.~Moneta\href
  {http://pos.sissa.it/archive/conferences/093/057/ACAT2010_057.pdf}{ {et~al.},
  ``The {R}oo{S}tats {P}roject'',} in \textit{ 13$^\text{th}$ International
  Workshop on Advanced Computing and Analysis Techniques in Physics Research
  (ACAT2010)}.
\newblock SISSA, 2010.
\newblock \href{http://www.arXiv.org/abs/1009.1003}{\texttt{arXiv:1009.1003}}.

\bibitem{Khachatryan:2014gga}
\hrefCMSnoop {}{{CMS Collaboration}, ``{Performance of the CMS missing
  transverse momentum reconstruction in pp data at $\sqrt{s} = 8\TeV$}'',}
  \textit{ JINST} \textbf{ 10} (2015) 02006,
  \href{http://dx.doi.org/10.1088/1748-0221/10/02/P02006}{\doi{10.1088/1748-0221/10/02/P02006}},
\href{http://www.arXiv.org/abs/1411.0511}{\texttt{arXiv:1411.0511}}.

\bibitem{CMS-PAS-LUM-17-001}
\href {https://cds.cern.ch/record/2257069}{{CMS Collaboration}, ``{CMS}
  luminosity measurements for the 2016 data taking period'',} CMS Physics
  Analysis Summary CMS-PAS-LUM-17-001, 2017.

\bibitem{Khachatryan:2014uva}
\hrefCMSnoop {}{{CMS Collaboration}, ``{Differential cross section measurements
  for the production of a W boson in association with jets in proton-proton
  collisions at $\sqrt s=7\TeV$}'',} \textit{ Phys. Lett. B} \textbf{ 741}
  (2015) 12,
  \href{http://dx.doi.org/10.1016/j.physletb.2014.12.003}{\doi{10.1016/j.physletb.2014.12.003}},
\href{http://www.arXiv.org/abs/1406.7533}{\texttt{arXiv:1406.7533}}.

\bibitem{Chatrchyan:2013uza}
\hrefCMSnoop {}{{CMS Collaboration}, ``{Measurement of the production cross
  section for a W boson and two b jets in pp collisions at $\sqrt{s}=7
  \TeV$}'',} \textit{ Phys. Lett. B} \textbf{ 735} (2014) 204,
  \href{http://dx.doi.org/10.1016/j.physletb.2014.06.041}{\doi{10.1016/j.physletb.2014.06.041}},
\href{http://www.arXiv.org/abs/1312.6608}{\texttt{arXiv:1312.6608}}.

\bibitem{Khachatryan:2014zya}
\hrefCMSnoop {}{{CMS Collaboration}, ``{Measurements of jet multiplicity and
  differential production cross sections of Z+jets events in proton-proton
  collisions at $\sqrt{s} = 7\TeV$}'',} \textit{ Phys. Rev. D} \textbf{ 91}
  (2015) 052008,
  \href{http://dx.doi.org/10.1103/PhysRevD.91.052008}{\doi{10.1103/PhysRevD.91.052008}},
\href{http://www.arXiv.org/abs/1408.3104}{\texttt{arXiv:1408.3104}}.

\bibitem{Chatrchyan:2014dha}
\hrefCMSnoop {}{{CMS Collaboration}, ``{Measurement of the production cross
  sections for a Z boson and one or more b jets in pp collisions at $\sqrt{s} =
  7\TeV$}'',} \textit{ JHEP} \textbf{ 06} (2014) 120,
  \href{http://dx.doi.org/10.1007/JHEP06(2014)120}{\doi{10.1007/JHEP06(2014)120}},
\href{http://www.arXiv.org/abs/1402.1521}{\texttt{arXiv:1402.1521}}.

\bibitem{Chatrchyan:1642680}
\hrefCMSnoop {}{{CMS Collaboration}, ``{Observation of the associated
  production of a single top quark and a $W$ boson in pp collisions at $\sqrt s
  = 8 \TeV$}'',} \textit{ Phys. Rev. Lett.} \textbf{ 112} (2014) 231802,
  \href{http://dx.doi.org/10.1103/PhysRevLett.112.231802}{\doi{10.1103/PhysRevLett.112.231802}},
  \href{http://www.arXiv.org/abs/1401.2942}{\texttt{arXiv:1401.2942}}.

\bibitem{Khachatryan:2016txa}
\hrefCMSnoop {}{{CMS Collaboration}, ``{Measurement of the ZZ production cross
  section and Z $\to \ell^+\ell^-\ell'^+\ell'^-$ branching fraction in pp
  collisions at $\sqrt s = 13\TeV$}'',} \textit{ Phys. Lett. B} \textbf{ 763}
  (2016) 280,
  \href{http://dx.doi.org/10.1016/j.physletb.2016.10.054}{\doi{10.1016/j.physletb.2016.10.054}},
  \href{http://www.arXiv.org/abs/1607.08834}{\texttt{arXiv:1607.08834}}.

\bibitem{Khachatryan:2016tgp}
\hrefCMSnoop {}{{CMS Collaboration}, ``{Measurement of the WZ production cross
  section in pp collisions at $\sqrt s = 13\TeV$}'',} \textit{ Phys. Lett. B}
  \textbf{ 766} (2017) 268,
  \href{http://dx.doi.org/10.1016/j.physletb.2017.01.011}{\doi{10.1016/j.physletb.2017.01.011}},
  \href{http://www.arXiv.org/abs/1607.06943}{\texttt{arXiv:1607.06943}}.

\bibitem{Heinemeyer:2013tqa}
\hrefCMSnoop {}{S.~Heinemeyer {et~al.}, ``Handbook of {LHC} {Higgs} cross
  sections: 3. {Higgs} properties'',} CERN Report CERN-2013-004, 2013.
\newblock
  \href{http://dx.doi.org/10.5170/CERN-2013-004}{\doi{10.5170/CERN-2013-004}},
  \href{http://www.arXiv.org/abs/1307.1347}{\texttt{arXiv:1307.1347}}.

\bibitem{bib:CLS1}
\hrefCMSnoop {}{A.~L. Read, ``{Presentation of search results: the CL$_s$
  technique}'',} \textit{ J. Phys. G} \textbf{ 28} (2002) 2693,
  \href{http://dx.doi.org/10.1088/0954-3899/28/10/313}{\doi{10.1088/0954-3899/28/10/313}}.

\bibitem{bib:CLS2}
\hrefCMSnoop {}{T.~Junk, ``{Confidence level computation for combining searches
  with small statistics}'',} \textit{ Nucl. Instrum. Meth. A} \textbf{ 434}
  (1999) 435,
  \href{http://dx.doi.org/10.1016/S0168-9002(99)00498-2}{\doi{10.1016/S0168-9002(99)00498-2}},
\href{http://www.arXiv.org/abs/hep-ex/9902006}{\texttt{arXiv:hep-ex/9902006}}.

\bibitem{bib:CLS3}
\hrefCMSnoop {}{G.~Cowan, K.~Cranmer, E.~Gross, and O.~Vitells, ``Asymptotic
  formulae for likelihood-based tests of new physics'',} \textit{ Eur. Phys. J.
  C} \textbf{ 71} (2011) 1554,
  \href{http://dx.doi.org/10.1140/epjc/s10052-011-1554-0}{\doi{10.1140/epjc/s10052-011-1554-0}},
  \href{http://www.arXiv.org/abs/1007.1727}{\texttt{arXiv:1007.1727}}.
[Erratum: \DOI{10.1140/epjc/s10052-013-2501-z}].

\bibitem{presentDM}
\hrefCMSnoop {}{A.~Boveia {et~al.}, ``Recommendations on presenting {LHC}
  searches for missing transverse energy signals using simplified $s$-channel
  models of dark matter'',} (2016).
  \href{http://www.arXiv.org/abs/1603.04156}{\texttt{arXiv:1603.04156}}.

\bibitem{CresstII}
\hrefCMSnoop {}{{{CRESST-II}} Collaboration, ``Results on light dark matter
  particles with a low-threshold {CRESST-II} detector'',} \textit{ Eur. Phys.
  J. C} \textbf{ 76} (2016) 25,
  \href{http://dx.doi.org/10.1140/epjc/s10052-016-3877-3}{\doi{10.1140/epjc/s10052-016-3877-3}},
  \href{http://www.arXiv.org/abs/1509.01515}{\texttt{arXiv:1509.01515}}.

\bibitem{CDMSLite}
\hrefCMSnoop {}{{SuperCDMS} Collaboration, ``New results from the search for
  low-mass weakly interacting massive particles with the {CDMS} low ionization
  threshold experiment'',} \textit{ Phys. Rev. Lett.} \textbf{ 116} (2016)
  071301,
  \href{http://dx.doi.org/10.1103/PhysRevLett.116.071301}{\doi{10.1103/PhysRevLett.116.071301}},
  \href{http://www.arXiv.org/abs/1509.02448}{\texttt{arXiv:1509.02448}}.

\bibitem{LUX}
\hrefCMSnoop {}{{LUX} Collaboration, ``Results from a search for dark matter in
  the complete {LUX} exposure'',} \textit{ Phys. Rev. Lett.} \textbf{ 118}
  (2017) 021303,
  \href{http://dx.doi.org/10.1103/PhysRevLett.118.021303}{\doi{10.1103/PhysRevLett.118.021303}},
  \href{http://www.arXiv.org/abs/1608.07648}{\texttt{arXiv:1608.07648}}.

\bibitem{XENON1T}
\hrefCMSnoop {}{{XENON} Collaboration, ``First dark matter search results from
  the {XENON1T} experiment'',} \textit{ Phys. Rev. Lett.} \textbf{ 119} (2017)
  181301,
  \href{http://dx.doi.org/10.1103/PhysRevLett.119.181301}{\doi{10.1103/PhysRevLett.119.181301}},
  \href{http://www.arXiv.org/abs/1705.06655}{\texttt{arXiv:1705.06655}}.

\bibitem{PandaxII}
\hrefCMSnoop {}{{PandaX-II} Collaboration, ``Dark matter results from
  54-ton-day exposure of {P}anda{X}-{II} experiment'',} \textit{ Phys. Rev.
  Lett.} \textbf{ 119} (2017) 181302,
  \href{http://dx.doi.org/10.1103/PhysRevLett.119.181302}{\doi{10.1103/PhysRevLett.119.181302}},
  \href{http://www.arXiv.org/abs/1708.06917}{\texttt{arXiv:1708.06917}}.

\bibitem{Jiang:2018pic}
\hrefCMSnoop {}{{CDEX} Collaboration, ``Limits on light weakly interacting
  massive particles from the first 102.8\,kg ${\times}$ day data of the
  {CDEX-10} experiment'',} \textit{ Phys. Rev. Lett.} \textbf{ 120} (2018)
  241301,
  \href{http://dx.doi.org/10.1103/PhysRevLett.120.241301}{\doi{10.1103/PhysRevLett.120.241301}},
\href{http://www.arXiv.org/abs/1802.09016}{\texttt{arXiv:1802.09016}}.

\end{thebibliography}\endgroup
